# Some generalizations of the tornado model

**S.N. Artekha**

*Space Research Institute of RAS, Moscow 117997, Russia*

Corresponding author. *E-mail address:* sergey.arteha@gmail.com (Sergey Artekha).

**ABSTRACT**

A possible mechanism of tornado generation is associated with the Schwarzschild criterion for convective instability. The supercooled cloud-air mixture near the axis starts to descend and, due to the super adiabatic temperature gradient, the overheated near-surface air mass rises. The complete system of equations for full model includes equations for the stream function, for the azimuthal velocity component, and for the change in density. The refined model considers rotating convection within the framework of ideal hydrodynamics. The simplest model uses a separate equation for the stream function and can be applied to the initial stage of vortex generation. Within the framework of the simplest model, an analytical solution is combined from 8 linear dependencies by height. Rotation increases due to the conservation of angular momentum. Within the framework of the refined model, some unidirectional differential rotation arises, but it depends on boundary conditions (can be local). The full model studies the tornado development as rotating convection from the very beginning to the quasi-stationary stage. Velocity components increase in a coordinated manner. Two cases are considered: a single whirlwind and a two-eyed tornado. Radial flows converge towards the central part of the tornado near the earth surface, and there is an outflow from the "wall" to the axis and to the periphery near the top of the tornado. The maximum vertical velocity component is achieved at a certain radial distance at a certain height. The maximum azimuthal velocity component is achieved in the tornado "wall". Tornado generation criteria are discussed.

## 1. Introduction

Atmospheric vortex processes significantly affect the weather. Among the vortex processes, the most noticeable are concentrated vortices, which represent spatially localized non-stationary vertically elongated vortex structures. Their characteristic vertical scale ranges from several hundred meters to several kilometers and characteristic transverse scale ranges from several meters to several hundred meters (sometimes up to several kilometers). The most intense concentrated vortices are tornado (Nalivkin, 1983) and whirlwind including waterspout (Church et al., 1979), firespout (Thorarinsson, Vonnegut, 1964; Tohidi et al., 2018). In these vortices, an upward spiral motion is observed with a maximum azimuthal velocity component at some distance from the axis of rotation (in the reference frame associated with the axis of rotation, all velocity components tend to zero at the periphery of the vortex). Some of the tornadoes can exceed 4 kilometers in diameter and travel hundreds of kilometers with wind speeds more than 500 km/h. A distinctive feature of a tornado is that it always comes into contact with the parent thundercloud at the top (Nalivkin, 1983; Grazulis, 2001; Ben-Amots, 2016). At the initial moment, the trunk of a tornado starts to descend from such a thundercloud. The air inside the funnel descends (eye of the tornado) and the air outside (wall of the tornado and nearby) rises, rotating rapidly.

Information on observations of whirlwinds and tornadoes, including some of their characteristics (sometimes there are some terminological disagreements among observers, physicists and meteorologists), can be found in scientific literature (see, for example, (AGU, 2018; Churc et al., 1993; Chernokulsky et al., 2020; Chernokulsky et al., 2021)), on the websites of specialized organizations in different countries and on popular Internet resources. The width of a tornado at the terrestrial surface ranges from 2 m to 4.2 km; the height usually ranges from 150 m to 1500 m, but can be more (if you take into account the rotating cloud, it can reach up to 10 km); wind speeds inside the tornado wall ranges from 100 km/h to 575 km/h; maximum azimuthal velocities are discovered at an altitude of 20-60 m near the ground; the translation speed of the tornado as a whole ranges from 0 to 120 km/h; the distance traveled ranges from 2 m to 352 km; the lifetime of a tornado ranges from several minutes to 3.5 hours; the time it takes for a tornado to form ranges from a few minutes to about an hour (excluding the preparatory stages, for example, the formation of the rotating supercell itself). The visible shape of tornadoes can be different: a narrow funnel, a “stovepipe” tornado, “wedge tornadoes”, narrow tubes or ropes, multiple-vortex tornadoes, etc. Average tornado characteristics vary noticeably depending on the season and geographic location of occurrence.

There are numerous field observations of the tornado phenomenon (Bluestein, 2005; Lee, Wurman, 2005; Wurman, Alexander, 2005; Kosiba, Wurman, 2010). Modern research methods also include numerical modeling (Lewellen, Lewellen, 2007; Pandey, 2025) and laboratory simulation of physical processes (Church, Snow, 1993; Lv et al., 2023). Many works have been devoted to the study of tornadoes, including the conditions of their generation, thermodynamics, the dynamics of their development, various physical models, and numerical modeling (see, for example, (Yih, 2007; Makarieva et al., 2011; Ben-Amots, 2016), as well as reviews (Serrin, 1972; Davies-Jones, 2015) and references therein). Generally speaking, there are many tornado models (analytical, empirical and numerical) and theories: supercell theory, rear-flank downdraft theory, multiple vortex theory, self-organization, electric theory etc. Many authors point to a downward flow in a funnel of tornado (Serrin, 1972; Davies-Jones, 2015; Ben-Amots, 2016). Review (Davies-Jones, 2015) draws attention to the rotation of the updraft during a turning shear, the need for the downdraft, the baroclinic and barotropic mechanisms, and the role of friction with the underlying surface. The role of condensation is emphasized in the works (Makarieva et al., 2011; Ben-Amots, 2016). The two-fluid model consists of the following. Expanding moist air from the cloud is adiabatically cooled in the funnel and condensation begins. Taking into account the displacement of droplets due to centrifugal force, the pressure is greatly reduced. A large pressure difference leads to an increase in flow speed, i.e. to intensifying tornadoes. Considering the balance of forces, some estimates were made in (Zhang et al., 2023) that viscosity can be neglected for calculating radial and vertical velocity components, but viscosity is important for determining the azimuthal velocity component.

Despite the vast capabilities of satellite sensing and supercomputers (Markowski, Richardson, 2014; Orf, 2019; Orf et al., 2017; Dixon et al., 2024), only 20 % of observed supercells become a tornado. Sometimes tornadoes appear not only from supercells, but from a squall line as well. The National Weather Service has a false alarm rate of about 70 %. Despite the noticeable progress in research of this phenomenon, many questions related to the conditions for the tornado generation, their dynamics, and the possibility of prediction and influence on them remain open (Markowski, Richardson, 2014).

Due to the wide variety of tornado characteristics, it is useful to have a simple but flexible analytical model. A hydrodynamic model of generation of an axially symmetric vortex structures in a convectively unstable atmosphere at the initial stage was proposed in Onishchenko et al., 2014 (the development of this model is considered in (Onishchenko et al., 2020; Horton et al., 2016)). An analytical model of generation of a spatially limited vortex, applicable for arbitrary values of the coordinates $r$ and $z$, was presented in (Onishchenko et al., 2023). Solutions represent some combinations from Bessel functions. A new analytical model for the generation of axisymmetric tornado-type vortices at the initial stage has been developed in (Artekha, 2024).

It is well known that finite-element analysis (broadly used for tornado-like structures) requires the solution of large linear systems that can be computationally intensive. Moreover, the initial and boundary conditions in a real problem cannot be known always and everywhere exactly: there are always stochastic processes, lost data, etc. Therefore, it is worth mentioning the latest trends in the development of research methods that may prove promising for the tasks under consideration. We are talking about Deep Learning, including the convolutional neural networks (Bolandi et al., 2022), the hierarchical deep-learning neural network (Zhang et al., 2021), the physics-informed deep learning (Xu et al., 2024). These techniques can generate results significantly faster than conventional run-time analysis, optimizing the nodal coordinates. For the last method, there is no need of boundary and initial conditions. It uses a mesh-free learning model and is not required to be solved in the whole computational domain as that in numerical methods. Such approaches can help determine criteria for tornadogenesis.

The goals of this work are the following:

– within the framework of the constructed simple ideal analytical model (Artekha, 2024), which includes a number of free parameters, to describe the structure of the tornado velocity field for nonlinear z-profiles;

– in the approximation of ideal hydrodynamics, to derive a refined system of equations and obtain their numerical solution that describes the developing instability at an initial stage in a stratified atmosphere;

– taking into account viscosity, to obtain numerical solutions, which describe the generation and development of a tornado localized in space at any stage.

The structure of the article is as follows. Section 2 presents the general system of equations that can manage developing instability. Section 3 considers various simplifications of the resulting system of equations in viscous and non-viscous cases. Section 4 discusses the structure of the tornado velocity field for linear and nonlinear z-profiles within the framework of a simplified model of ideal hydrodynamics. Section 5 describes numerical solutions of the refined tornado model. Section 6 presents numerical solutions of the complete system of equations with the transition to a quasi-stationary stage. Section 7 discusses the main findings of the study and contains conclusions.

## 2. System of basic equations

We will describe the tornado generation in a cylindrical coordinate system $(r, \varphi, z)$ with the $z$ axis directed vertically upward. In axisymmetric case, all quantities are independent of the angle of the cylindrical coordinate system $\varphi$, i.e. $\partial/\partial\varphi = 0$. The values of pressure and density, respectively, can be written by the sum of the equilibrium (initial) and perturbed values:

$$p = p_0(z) + \tilde{p}(t, r, z),\ \rho = \rho_0(z) + \tilde{\rho}(t, r, z). \tag{1}$$

At the initial moment, the condition of hydrostatic equilibrium

$$\frac{dp_0}{dz} = -\rho_0(z) g \tag{2}$$

determines the altitude dependence of the equilibrium pressure:

$$p_0(z) = p_{00}\, exp\left(-\frac{z}{H}\right),\ \ H = \frac{c_s^2}{\gamma_a g},\ \ c_s = \sqrt{\frac{\gamma_a p_{00}}{\rho_{00}}}, \tag{3}$$

where $H$ is the local scale of atmospheric height, $c_s$ is the speed of sound, $\boldsymbol{g} = -g\hat{\boldsymbol{z}}$ is the gravitational acceleration ($\hat{\boldsymbol{z}}$ is the unit vector along the vertical axis), and $\gamma_a$ is the adiabatic exponent (the ratio of specific heat capacities $\frac{c_P}{c_V}$); $p_{00}$ and $\rho_{00}$, respectively, are the initial values of pressure and density near the earth's surface at $z = 0$.

The flow velocity $\boldsymbol{v} = (v_r, v_\varphi, v_z)$ represents the sum of the poloidal component $\boldsymbol{v}_p$ and the azimuthal component $v_\varphi \hat{\boldsymbol{e}}_\varphi$, i.e. $\boldsymbol{v} = \boldsymbol{v}_p + v_\varphi \hat{\boldsymbol{e}}_\varphi$, where $\hat{\boldsymbol{e}}_\varphi$ is the corresponding unit vector. The original system of equations consists of the following equations:

1) continuity equation:

$$\frac{\partial \rho}{\partial t} = -\nabla \cdot (\rho \boldsymbol{v}) = -\rho(\nabla \cdot \boldsymbol{v}) - (\boldsymbol{v} \cdot \nabla)\rho. \tag{4}$$

We consider the emerging flow to be incompressible:

$$\frac{1}{r}\frac{\partial (r v_r)}{\partial r} + \frac{\partial v_z}{\partial z} = 0, \tag{5}$$

then the distribution of the perturbed density will be determined by the equation:

$$\frac{\partial \tilde{\rho}}{\partial t} = -v_r \frac{\partial \tilde{\rho}}{\partial r} - v_z \frac{\partial (\rho_0 + \tilde{\rho})}{\partial z}. \tag{4*}$$

2) Navier-Stokes equations:

$$\frac{\partial v_r}{\partial t} + v_r \frac{\partial v_r}{\partial r} + v_z \frac{\partial v_r}{\partial z} - \frac{v_\varphi^2}{r} = -\frac{1}{\rho}\frac{\partial p}{\partial r} + \nu\left(\frac{1}{r}\frac{\partial}{\partial r}\left(r\frac{\partial v_r}{\partial r}\right) + \frac{\partial^2 v_r}{\partial z^2} - \frac{v_r}{r^2}\right), \tag{6}$$

$$\frac{\partial v_z}{\partial t} + v_r \frac{\partial v_z}{\partial r} + v_z \frac{\partial v_z}{\partial z} = -\frac{1}{\rho}\frac{\partial p}{\partial z} + \nu\left(\frac{1}{r}\frac{\partial}{\partial r}\left(r\frac{\partial v_z}{\partial r}\right) + \frac{\partial^2 v_z}{\partial z^2}\right), \tag{7}$$

$$\frac{\partial v_\varphi}{\partial t} + v_r \frac{\partial v_\varphi}{\partial r} + v_z \frac{\partial v_\varphi}{\partial z} + \frac{v_r v_\varphi}{r} = \nu\left(\frac{1}{r}\frac{\partial}{\partial r}\left(r\frac{\partial v_\varphi}{\partial r}\right) + \frac{\partial^2 v_\varphi}{\partial z^2} - \frac{v_\varphi}{r^2}\right). \tag{8}$$

3) Equation of state of an ideal gas:

$$\frac{p}{\rho T} = const, \tag{9}$$

$T$ is the temperature.

4) The transport equation (as Euler's derivative) for potential temperature $\theta = p^{1/\gamma_a}/\rho$:

$$\frac{d\theta}{dt} = 0. \tag{10}$$

The instability of the stratified atmosphere leads to the generation of vertical currents (thermodynamic model). It is well known that square of the Brunt–Väisälä frequency or buoyancy frequency

$$\omega_g^2 \equiv \frac{g}{\theta}\frac{d\theta}{dz} = g\left(\frac{\gamma_a - 1}{\gamma_a H} + \frac{1}{T}\frac{dT}{dz}\right) \tag{11}$$

characterizes internal gravity waves (IGW). In the case of $\omega_g^2 > 0$, convective instability does not arise. In this case, any disturbances are carried away from the region of their origin by the IGWs, and the atmosphere is stable. The atmosphere is unstable stratified if $\omega_g^2 < 0$. Due to solar heating of the terrestrial surface, the vertical temperature gradient d$T$/d$z$ (the second term of the Brunt–Väisälä frequency in (11)) is negative, and its absolute value can exceed the first term in (11). The Schwarzschild criterion for convective instability is:

$$-\frac{1}{T}\frac{dT}{dz} > \rho_0 g \frac{\gamma_a - 1}{\gamma_a p_0}.$$

In such a case, disturbances are not carried away from the region of their origin by IGWs. As a result, in the instability region, disturbances can turn into unstable growing structures.

We will consider the problem from the most complete system of equations to the simplest system of equations. When simplifying the basic system of equations, we will follow the procedure developed in (Renno et al., 1998; Onishchenko et al., 2014; Onishchenko et al., 2015). For all three simplifications, we are only interested in the case of $\omega_g^2 < 0$.

## 3. Some simplified systems of equations

To satisfy the equation (5), we will use the stream function in cylindrical coordinates. Poloidal velocity components are obtained from the stream function $\psi(t, r, \varphi, z)$ as follows:

$$v_r = -\frac{1}{r}\frac{\partial\psi}{\partial z}, \quad v_z = \frac{1}{r}\frac{\partial\psi}{\partial r}. \tag{12}$$

The equation (4*) can be re-written in terms of $\psi(t, r, \varphi, z)$ and $\chi = \frac{\tilde{\rho}}{\rho_0}$ as follows:

$$r\frac{\partial\chi}{\partial t} + J(\psi, \chi) + (1 + \chi)\frac{\partial\psi}{\partial r}\frac{d\ln\rho_0}{dz} = 0, \tag{13}$$

where $J(a,b) = \frac{\partial a}{\partial r}\frac{\partial b}{\partial z} - \frac{\partial a}{\partial z}\frac{\partial b}{\partial r}$ is the Jacobian determinant, $\rho_0(z) = \rho_{00}(1 - l_0 z/T)^{\frac{g\mu}{R_0 l_0} - 1}$, $g \sim 9.8\ \mathrm{m/c^2}$, $l_o \sim 6$ K/km, $R_0 = 8.31446\ \mathrm{J}\ K^{-1}\mathrm{mol}^{-1}$, $\mu \sim 29$ g/mol.

The equation (8) for the azimuthal velocity component can be re-written as follows:

$$\frac{\partial v_\varphi}{\partial t} - \frac{1}{r}\frac{\partial \psi}{\partial z}\frac{\partial v_\varphi}{\partial r} + \frac{1}{r}\frac{\partial \psi}{\partial r}\frac{\partial v_\varphi}{\partial z} - \frac{\partial \psi}{\partial z}\frac{v_\varphi}{r^2} = \nu\left(\frac{1}{r}\frac{\partial}{\partial r}\left(r\frac{\partial v_\varphi}{\partial r}\right) + \frac{\partial^2 v_\varphi}{\partial z^2} - \frac{v_\varphi}{r^2}\right). \quad (14)$$

Using equations (6) and (7), we can exclude the value of $p$ and find the $\varphi$-component of the curl of the velocity. As a result, we obtain the following equation:

$$\frac{\partial}{\partial t}\Delta^*\psi + \frac{1}{r}J(\psi, \Delta^*\psi) + \frac{2}{r^2}\frac{\partial \psi}{\partial z}\Delta^*\psi + 2v_\varphi\frac{\partial v_\varphi}{\partial z} + \nu r\Delta\left(\frac{1}{r}\Delta^*\psi\right) + \frac{\nu}{r^2}\frac{\partial^2 \psi}{\partial z^2} + gr\frac{\partial \ln\rho}{\partial r} +$$

$$+\frac{\partial \ln\rho}{\partial r}\left(\frac{\partial}{\partial t}\frac{\partial \psi}{\partial r} + \frac{1}{r}\frac{\partial \psi}{\partial r}\frac{\partial^2 \psi}{\partial z\partial r} - \frac{\partial \psi}{\partial z}\frac{\partial}{\partial r}\left(\frac{1}{r}\frac{\partial \psi}{\partial r}\right)\right) +$$

$$+\frac{\partial \ln\rho}{\partial z}\left(\frac{\partial}{\partial t}\frac{\partial \psi}{\partial z} + \frac{1}{r}\frac{\partial \psi}{\partial r}\frac{\partial^2 \psi}{\partial z^2} + v_\varphi^2 - \frac{\partial \psi}{\partial z}\frac{\partial}{\partial r}\left(\frac{1}{r}\frac{\partial \psi}{\partial z}\right)\right) = 0, \quad (15)$$

where the Laplace operator $\Delta$ and the operator $\Delta^*$, respectively, are defined as

$$\Delta = \frac{1}{r}\frac{\partial}{\partial r}\left(r\frac{\partial}{\partial r}\right) + \frac{\partial^2}{\partial z^2}, \quad (16)$$

$$\Delta^* = r\frac{\partial}{\partial r}\left(\frac{1}{r}\frac{\partial}{\partial r}\right) + \frac{\partial^2}{\partial z^2}. \quad (17)$$

The most complete system of equations (13)-(15) is derived without any simplifications. This system can be solved by numerical methods only. It can determine the values of $\psi, v_\varphi, \rho$ at arbitrary stage of the vortex development.

Within the framework of ideal hydrodynamics, neglecting small nonlinear terms, we can write a simpler system of equations describing the quantities $\psi, v_\varphi, \rho$ at the stage of vortex intensification. Instead of equations (13)-(15), respectively, we have the following equations:

$$r\frac{\partial \tilde{\chi}}{\partial t} + J(\psi, \tilde{\chi}) - \omega_g^2\frac{\partial \psi}{\partial r} = 0, \quad (18)$$

$$\frac{\partial v_\varphi}{\partial t} - \frac{1}{r}\frac{\partial \psi}{\partial z}\frac{\partial v_\varphi}{\partial r} + \frac{1}{r}\frac{\partial \psi}{\partial r}\frac{\partial v_\varphi}{\partial z} - \frac{\partial \psi}{\partial z}\frac{v_\varphi}{r^2} = 0, \quad (19)$$

$$\left(\frac{\partial^2}{\partial t^2} + \omega_g^2\right)\Delta^*\psi + \frac{d\ln\rho_0}{dz}\frac{\partial}{\partial t}\left(\frac{\partial^2\psi}{\partial z\partial t} + v_\varphi^2\right) + \frac{\partial}{\partial t}\left(\frac{1}{r}J(\psi, \Delta^*\psi) + \frac{2}{r^2}\frac{\partial \psi}{\partial z}\Delta^*\psi + 2v_\varphi\frac{\partial v_\varphi}{\partial z}\right) = 0, \quad (20)$$

where $\tilde{\chi} = g\frac{\tilde{\rho}}{\rho_0}$. The simpler system of equation (18)-(20) can also be solved by numerical methods. It can determine the quantities $\psi, v_\varphi, \rho$ at the stage of vortex intensification.

Finally, within the framework of ideal hydrodynamics, one can obtain one separate equation for the stream function, which describes the evolution of nonlinear internal gravity waves for $\omega_g^2 > 0$, and vortex generation at initial stage for $\omega_g^2 < 0$. According to (Onishchenko et al., 2014; Onishchenko et al., 2015), such an equation can be derived from (13) and (15) with the following simplifications:

$$|\tilde{\rho}| \ll \rho_0, |\tilde{p}| \ll p_0, \frac{\tilde{p}}{p_0} < \frac{\tilde{\rho}}{\rho_0}, \tilde{\chi} \sim \psi, v_z \gg v_r, \frac{\partial}{\partial r} \gg \frac{\partial}{\partial z},$$

$$L < H, \frac{\partial}{\partial r} \gg \frac{1}{L_\rho}, L_\rho^{-1} = \frac{d \ln \rho}{dz}, \frac{\omega^2}{c_s^2 k_r^2} L_\rho \frac{\partial}{\partial z} \ll 1,$$

here the last inequality assumes low frequency processes.

As a result, we have the following equation for the stream function:

$$\frac{\partial^2 \Delta_r^* \psi}{\partial t^2} + \omega_g^2 \Delta_r^* \psi + \frac{1}{r} \frac{\partial}{\partial t} J(\psi, \Delta_r^* \psi) = 0, \tag{21}$$

where $\Delta_r^*$ is the radial part of the operator (17). The Jacobian determinant represents the so-called vector nonlinearity. The simplest equation (21) allows us to find the stream function $\psi$ and, therefore, the radial and vertical components of the velocity. The change in density is given by equation (18), and the azimuthal component of velocity is given by equation (19).

Next, we will discuss the solutions that follow from the resulting systems of equations for all three cases, starting with the simplest one. To make it easier to navigate through mathematical derivations, we add a diagram illustrating the hierarchy of models. The sequence of mathematical operations for a tornado with one eye and with two eyes are similar (the method of selecting parameters for these cases and joining at boundaries is explained in detail in papers (Onishchenko et al., 2023) and (Artekha, 2024), respectively).

**The simplest analytical model**

Find a piecewise continuous stream function $\psi$ from (21) with $\left(\psi, \frac{\partial \psi}{\partial r}\right) \to 0$, when $r \to 0$ or $r \to \infty$

⇓

Select parameters; met at boundaries $\left(\psi, \frac{\partial \psi}{\partial r}\right)_i = \left(\psi, \frac{\partial \psi}{\partial r}\right)_{i+1}$

⇓

(12) ⇒ Find the poloidal velocity components $v_r, v_z$

⇓

(19) ⇒ Find the azimuthal velocity component $v_\varphi$
linear $f(z)$ ⇒ quadratures | nonlinear ⇒ numerically

⇓

Draw the velocity field

For the other two models the diagrams are similar to each other.

**The refined model**

Select initial and boundary conditions
(values from the simplest model)

⇓

Numerically solve the system of NDE (18)-(20)
(without viscosity)

⇓

Draw the velocity field

**The full model**

Select initial and boundary conditions

⇓

Numerically solve the complete system of NDE
(13)-(15) (viscous case)

⇓

Draw the velocity field and density graph

**4. Some solutions for the simplest tornado model**

We will consider the unstable case with $-\omega_g^2 \equiv \gamma^2 > 0$. Equation (21) can easily be solved by the method of separation of variables. To do this, we note that the sum of the first two terms can be zeroed out due to a time dependence of $\psi$, and the last term vanishes due to a specific radial dependence of $\psi$. From two such temporary solutions $\sim \exp(\pm\gamma t)$, we will combine the function $\sinh(\gamma t)$ so that the vortex generation starts at the initial time moment $t = 0$. Therefore, the stream function can be written in the form:

$$\psi(t, r, z) = v_0 r^2 f(z/L) \sinh(\gamma t)\, \Psi(r/r_0), \tag{22}$$

where $v_0 = const$ is some characteristic poloidal velocity, $L = const$ is the characteristic vertical spatial scale of the vortex, $r_0$ is the characteristic radius of the vortex, $f(z/L)$ is an arbitrary function.

The following linear differential equation for the function $\Psi(R)$ is derived in (Artekha, 2024; Onishchenko et al., 2023):

$$R^2 \frac{d^2\Psi}{dR^2} + 3R\frac{d\Psi}{dR} = \pm a_0^2 R^2 \Psi. \tag{23}$$

Its general solutions are:

(I) An arbitrary combination of functions $\Psi(R) = J_1(a_0 R)/R$ and $\Psi(R) = Y_1(a_0 R)/R$ for the negative sign before $a_0^2$ in (23);

(II) One of the real functions $\Psi(R) = I_1(a_0R)/R$ or $\Psi(R) = K_1(a_0R)/R$, for the positive sign before $a_0^2$ in (23);

(III) $\Psi(R) = C_0 + C'/R^2$ for a zero value of $a_0$.

However, since the equation (21) is nonlinear, the sum of different solutions is not a solution (constants in the right-hand side of (23) are different). Therefore, the areas of applicability of the solutions (I)-(III) with respect to the variable $R = r/r_0$ must not overlap. We can compose a single continuous solution from the piecewise exact solutions found above, as from blocks (by *r* and by *z*), so that it satisfies all the physical conditions on the axis and on the periphery of the tornado (often such a block structure of the vortex is observed when measuring its velocity field).

The following conditions are satisfied at the domain boundary for the simplest model:

1) in the center of the tornado (at $r = 0$): $v_r = 0, v_\varphi = 0$;

2) at the lower and upper boundaries of the tornado (at $z = 0$ and at $z = L$): $v_z = 0$;

3) on the periphery of the vortex: $v_r \to 0, v_z \to 0, v_\varphi \to 0$ for $r \to \infty$ ($r >> r_0$).

The localization of the vortex, the finiteness and continuity of the velocity impose restrictions on the stream function: for a real model, it must be finite and localized in the vertical and radial directions; particularly, $\left(\psi, \frac{\partial\psi}{\partial r}\right) \to 0$, when $r \to 0$ and $r \to \infty$. This means the requirements for regularity on the axis of symmetry of the tornado and for a sufficiently rapid decrease over large distances. At the boundary of blocks (*i*; *i*+1), the continuity conditions are met: $\left(\psi, \frac{\partial\psi}{\partial r}\right)_i = \left(\psi, \frac{\partial\psi}{\partial r}\right)_{i+1}$. The function $f(Z)$ of the dimensionless parameter $Z$ determines the dependence of radial and vertical velocity components on height (boundary values, location of maxima etc.).

A single (by radius $R = r/r_0$) vortex without an eye is considered in (Onishchenko et al., 2023), which can be applied to whirlwinds (including waterspout and fire spout). In this case the solution for poloidal velocity was chosen in the form (here $Z = z/L$):

$$v_r^{int} = -v_0 \frac{r_0}{L} f'(Z) \sinh(\gamma t) \frac{J_1(\delta_0 R)}{J_1(\delta_0)}, \tag{24}$$

$$v_r^{ext} = -v_0 \frac{r_0}{L} f'(Z) \sinh(\gamma t)\, m \frac{K_1(\delta R)}{K_1(\delta)}, \tag{25}$$

$$v_z^{int} = v_0 f(z/L) \sinh(\gamma t)\, \delta_0 \frac{J_0(\delta_0 R)}{J_1(\delta_0)}, \tag{26}$$

$$v_z^{ext} = -v_0 f(z/L) \sinh(\gamma t)\, m\delta \frac{K_0(\delta R)}{K_1(\delta)}. \tag{27}$$

Constants $\delta$, $\delta_0$ determine the profile of the velocity field of the vortex, and *m* ensures the smoothness of the function for all *R* (see Onishchenko et al., 2023).

If an eye (or several eyes) is observed in a tornado, its structure can be combined by radius ($R = r/r_0$) from several blocks. Possible structures of movements and scenarios for the development of a tornado and its dynamics are discussed in (Artekha, 2024). In this case the solution for poloidal velocity was found in the form (here $Z = z/L$):

$$v_r^{int} = v_0 \frac{r_0}{L} f'(Z) \sinh(\gamma t)\, C_0 J_1(a_0 R), \tag{28}$$

$$v_r^{centr} = -v_0 \frac{r_0}{L} f'(Z) \sinh(\gamma t)\, [C_1 J_1(a_1 R) + C_2 Y_1(a_1 R)], \tag{29}$$

$$v_r^{ext} = -v_0 \frac{r_0}{L} f'(Z) \sinh(\gamma t)\, C_3 K_1(a_2 R). \tag{30}$$

$$v_z^{int} = -v_0 f(Z) \sinh(\gamma t)\, C_0 a_0 J_0(a_0 R), \tag{31}$$

$$v_z^{centr} = v_0 f(Z) \sinh(\gamma t)\, a_1 [C_1 J_0(a_1 R) + C_2 Y_0(a_1 R)], \tag{32}$$

$$v_z^{ext} = -v_0 f(Z) \sinh(\gamma t)\, C_3 a_2 K_0(a_2 R). \tag{33}$$

All constants determine the profile of the velocity field of the tornado ensure the continuity and smoothness of these functions. The algorithm for finding them is described in detail in (Artekha, 2024).

The azimuthal component of velocity is determined by equation (19). In order for the solution $v_\varphi$ to be found by the method of separation of variables

$$v_\varphi = y(t) f_0(z/L) V_{\varphi r}(R), \tag{34}$$

$$y(t) = \exp\{c_0(\cosh(\gamma t) - 1)\}, \tag{35}$$

the function $f(z/L)$ must be linear in *z*. For two blocks in *z*, we write:

$$f(Z) = \begin{cases} k_1 Z, & 0 \le Z \le Z_1; \\ \frac{k_1 Z_1}{1 - Z_1}(1 - Z), & Z_1 < Z \le 1, \end{cases} \tag{36}$$

where $Z_1$ is the height where the maximum is reached. In this case, a power dependence for $f_0(Z)$ can be written:

$$f_0(Z) = \begin{cases} k_2 Z^n, & 0 \le Z \le Z_1; \\ \frac{k_2 Z_1^n}{(1 - Z_1)^n}(1 - Z)^n, & Z_1 < Z \le 1, \end{cases} \tag{37}$$

where $n > 0$ – any positive real number. The solution for the radial dependence of the azimuthal velocity component can be obtained in quadrature for each area *i*:

$$V_{\varphi r}^i(R) = C^i \exp\left\{\int_1^R \frac{-\alpha^i x - n x \hat{V}_z^i(x) - \hat{V}_r^i(x)}{x \hat{V}_r^i(x)} d\,x\right\}, \tag{38}$$

where $\hat{V}_r^i(R) = \tilde{v}_r^i/r_0, \hat{V}_z^i(R) = \tilde{v}_z^i/L$, and velocity components with the tilde icon mean the factor of the same function (24)–(27), or (28)-(33), but without temporary dependence and $z$-dependence.

The obtained exact solution of simplest model is applicable (on the periphery in $z$ and $r$) until the velocity decreases to the background values. The range of applicability of the model is limited by the region of convective instability and the region with non-zero torque. The analytical solution describes the initial stage of tornado generation.

We can construct a tornado from more than two parts in $z$ (by analogy with (36) and (37)). As an example, we consider a tornado with two eyes (Wurman et al., 1996; Bluestein, Pazmany, 1999; Wurman, Gill, 2000; Kosiba et al., 2008; Wurman, Kosiba, 2013). We will use the same constants as in (Artekha, 2024):

$$a_0 = 9.2228, a_1 = 2.5619, C_0 = 0.9611, C_1 = 1.6936, C_2 = 1.0724, R_1 = 0.4155, \quad (39)$$

$$a_2 = 4.0266, C_3 = 337.4226, R_2 = 1.5, \quad (40)$$

where $R_1 = r_1/r_0$ and $R_2 = r_2/r_0$ are the boundaries on which continuous and smooth joining of piecewise continuous solutions occurs. We take $n$ = 2, $L$ = 2500, $r_0 = 250$. Since the real range of tornado sizes is very wide, the choice of characteristic values of $L$ and $r_0$ is quite arbitrary. For a tornado with two eyes, we chose such rather larger sizes (for convenience of presentation), and for a single-cell tornado, we will choose smaller characteristic sizes of $L$ and $r_0$. Further, for convenience of comparison, we will perform all calculations with such selected values of $L$ and $r_0$. The radial profile of the normalized azimuthal velocity has the form shown in Fig. 1.

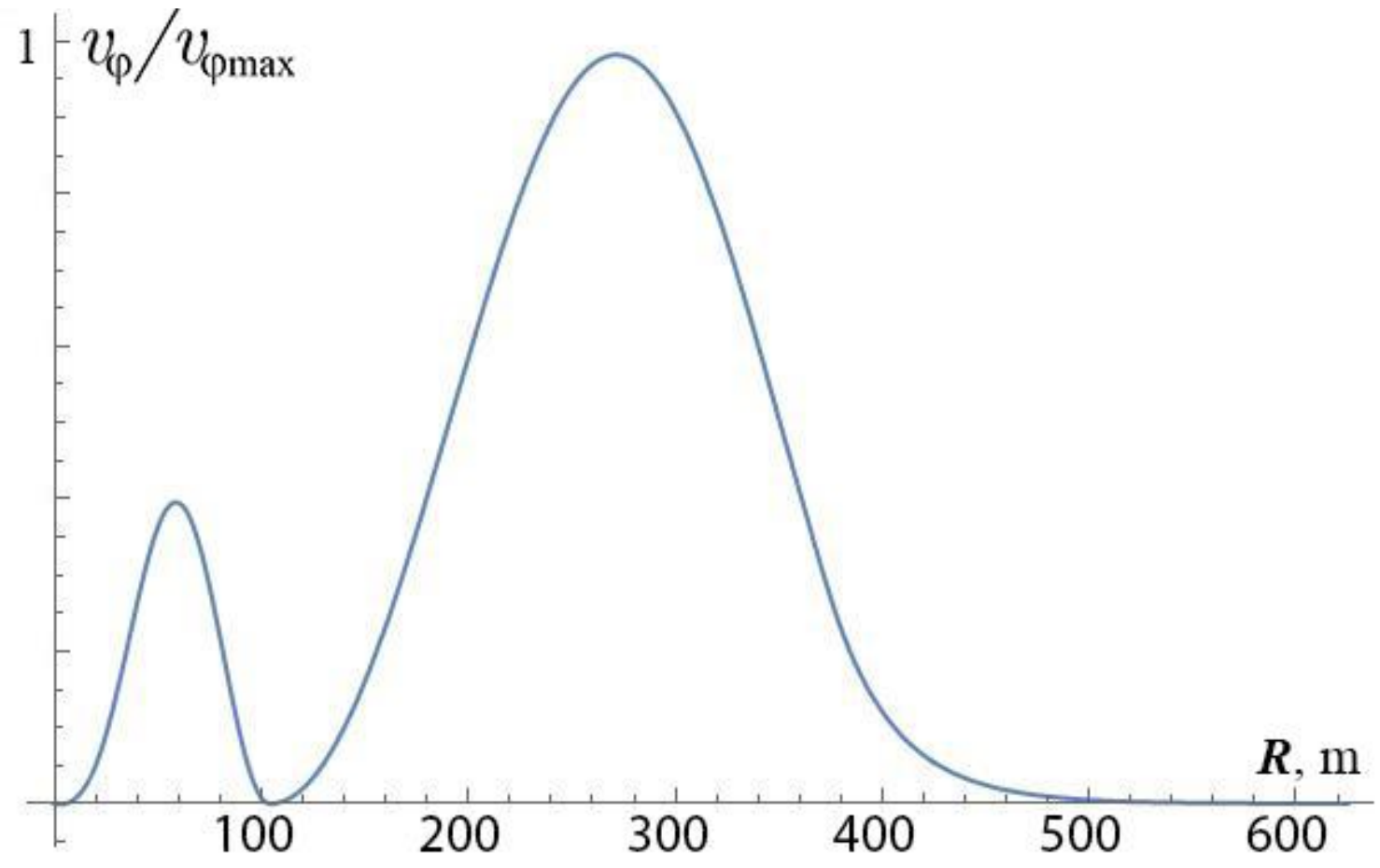

**FIG. 1.** The radial profile of the normalized azimuthal velocity component.

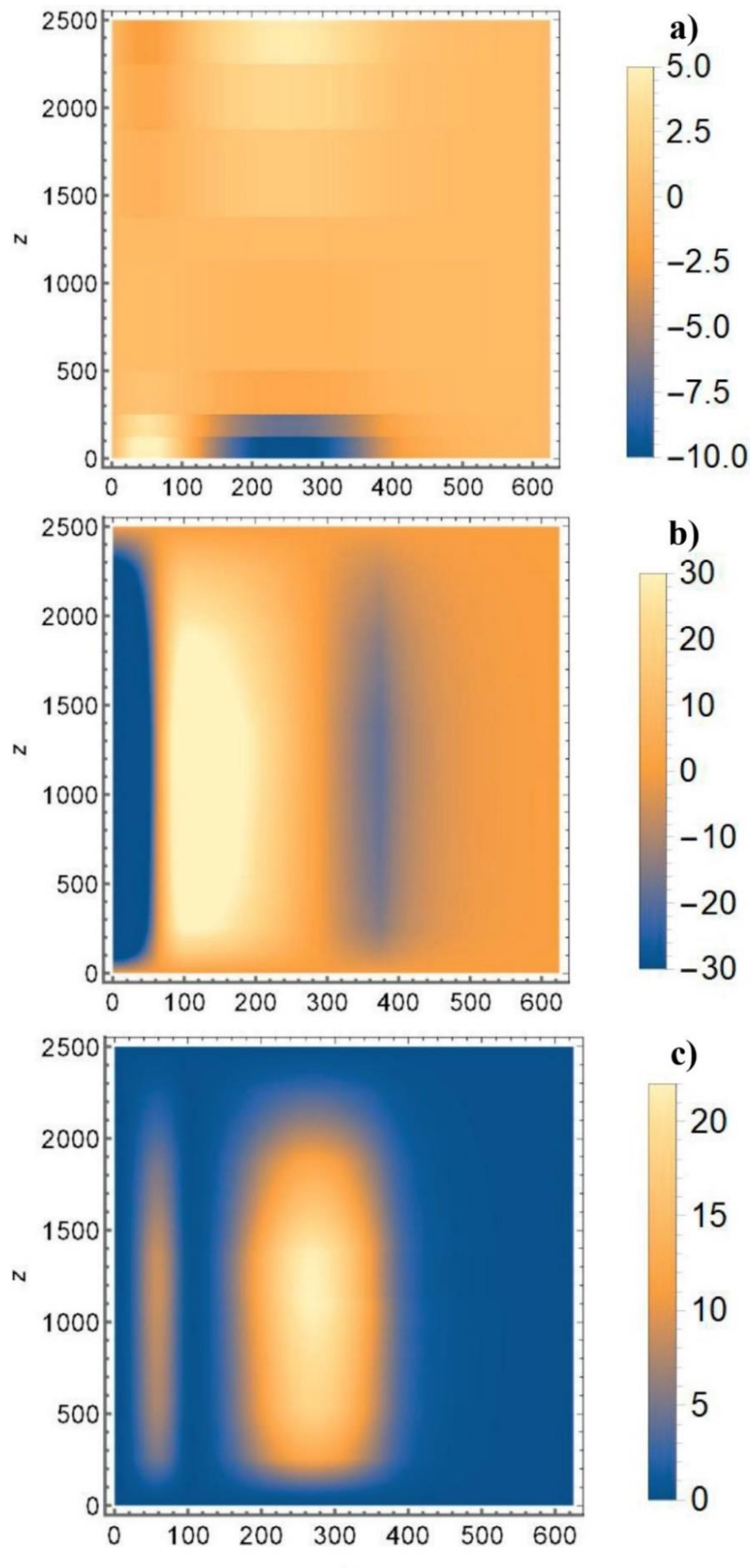

**FIG. 2.** The tornado velocity field in m/s depending on the coordinates $r$ and $z$ in meters for the linear profiles of the function $f(z/L)$: a) $v_r(r,z)$; b) $v_z(r,z)$; c) $v_\varphi(r,z)$.

For example, for the function $f(z/L)$, we will select 8 linear sections in $z/L$: (0, 0.05), (0.05, 0.1), (0.1, 0.2), (0.2, 0.45), (0.45, 0.55), (0.55, 0.75), (0.75, 0.9), (0.9, 1), with corresponding slope tangents: 9, 6, 1.5, 0.4, -0.1, -1.25, -2.4, -3.8. As a result, for the initial stage of tornado generation, we obtained the velocity distribution in Fig. 2. On the $r-z$ plane, the first graph depicts the radial velocity component, the second graph shows the vertical component, and the third graph demonstrates the azimuthal velocity component. We see the spatial localization of flows inside a tornado with extrema at certain distances $r$ and $z$. All movements disappear at the periphery. In this case, the velocity field and the order of all quantities resembles what is observed for the real tornadoes (Lee, Wurman, 2005; Kosiba et al., 2008; Kosiba, Wurman, 2010).

However, if the function $f(z/L)$ is nonlinear, the dependencies of azimuthal velocity on variables $z$ and $r$ in equation (19) cannot be separated, and this equation should be solved numerically. We will consider mono-whirlwind with one velocity extremum, analogous to case from (Onishchenko et al., 2023). All features of whirlwinds and tornadoes vary over very wide ranges. For universality, it is rather convenient to present mathematical solutions graphically in dimensionless variables (as a ratio to some analogous characteristic dimensional value: $R = r/r_0, Z = z/L, V = v/v_0$ etc.). In many cases the solution quickly becomes turbulent. For definiteness, we take

$$\delta_0 = 1.8412,\ r_1 = 1.7377,\ \delta = 2.5,\ m = 3.9363,\ r_0 = 0.1L. \tag{41}$$

First, we will choose the following function:

$$f(z/L) = \begin{cases} \tanh(10z/L), & 0 < z < L/2, \\ \exp\{-2(z/L-0.5)^2\}, & \mathrm{L}/2 < z < L, \end{cases} \tag{42}$$

and plot the dependence of the azimuthal velocity component on dimensionless variables $R$ and $Z$ (see Fig. 3).

If we choose $L = 800$ m and another function:

$$f(z/L) = \tanh\{20z(L-z)/L^2\}, \tag{43}$$

then we obtain another dependence of the azimuthal component of the velocity on real coordinates in meters, shown in Fig. 4. As we see in Figs. 3 and 4, many regions with multidirectional azimuthal velocities arise. Taking viscosity into account could lead to a preferential direction of rotation of such a turbulent structure. Perhaps, in such cases, it is necessary to search for a numerical solution of equation (19) simultaneously for all variables, including time.

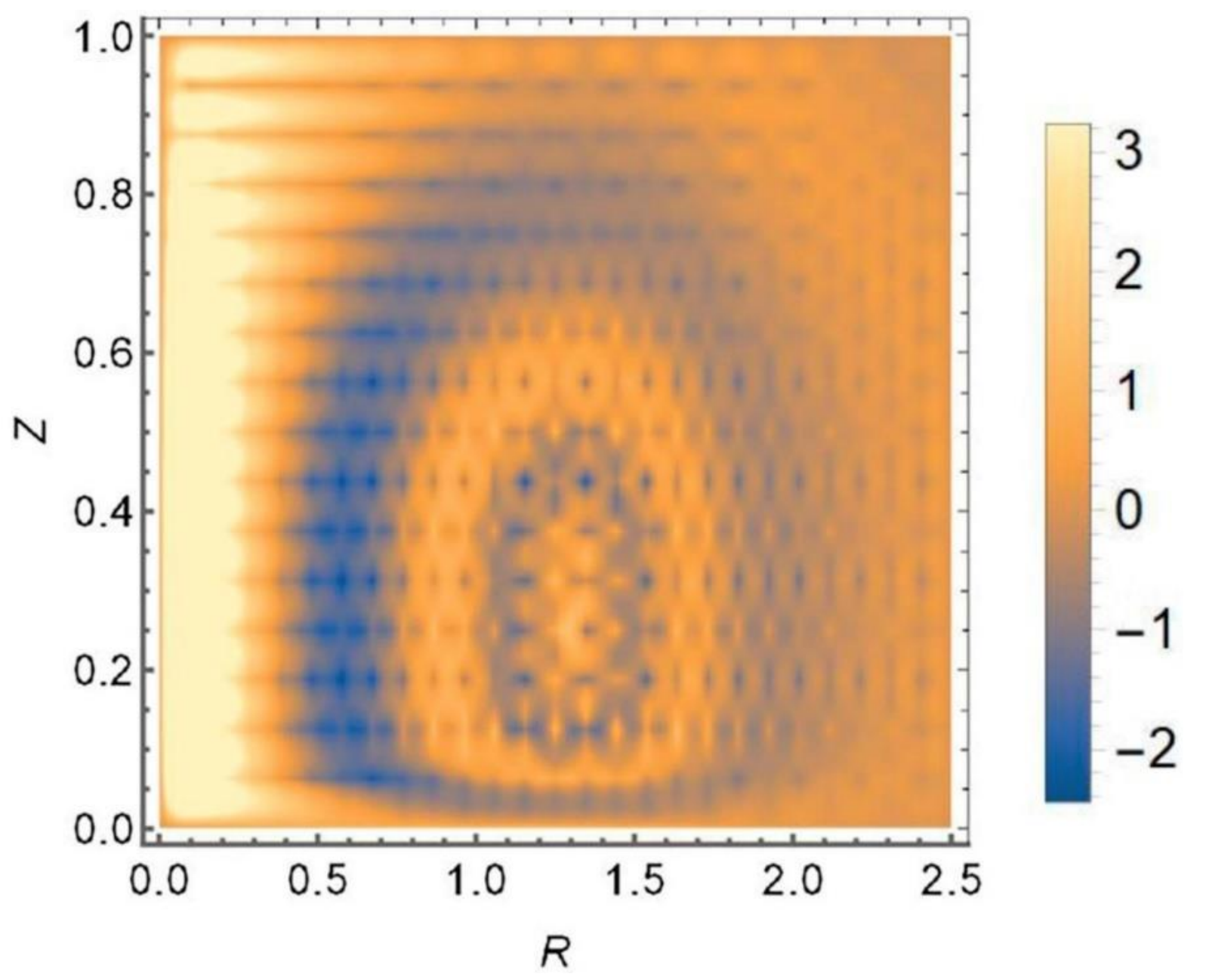

**FIG. 3.** The dependence of the dimensionless azimuthal component of whirlwind velocity on the dimensionless coordinates $R$ and $Z$ for the function $f(z/L)$ from (42).

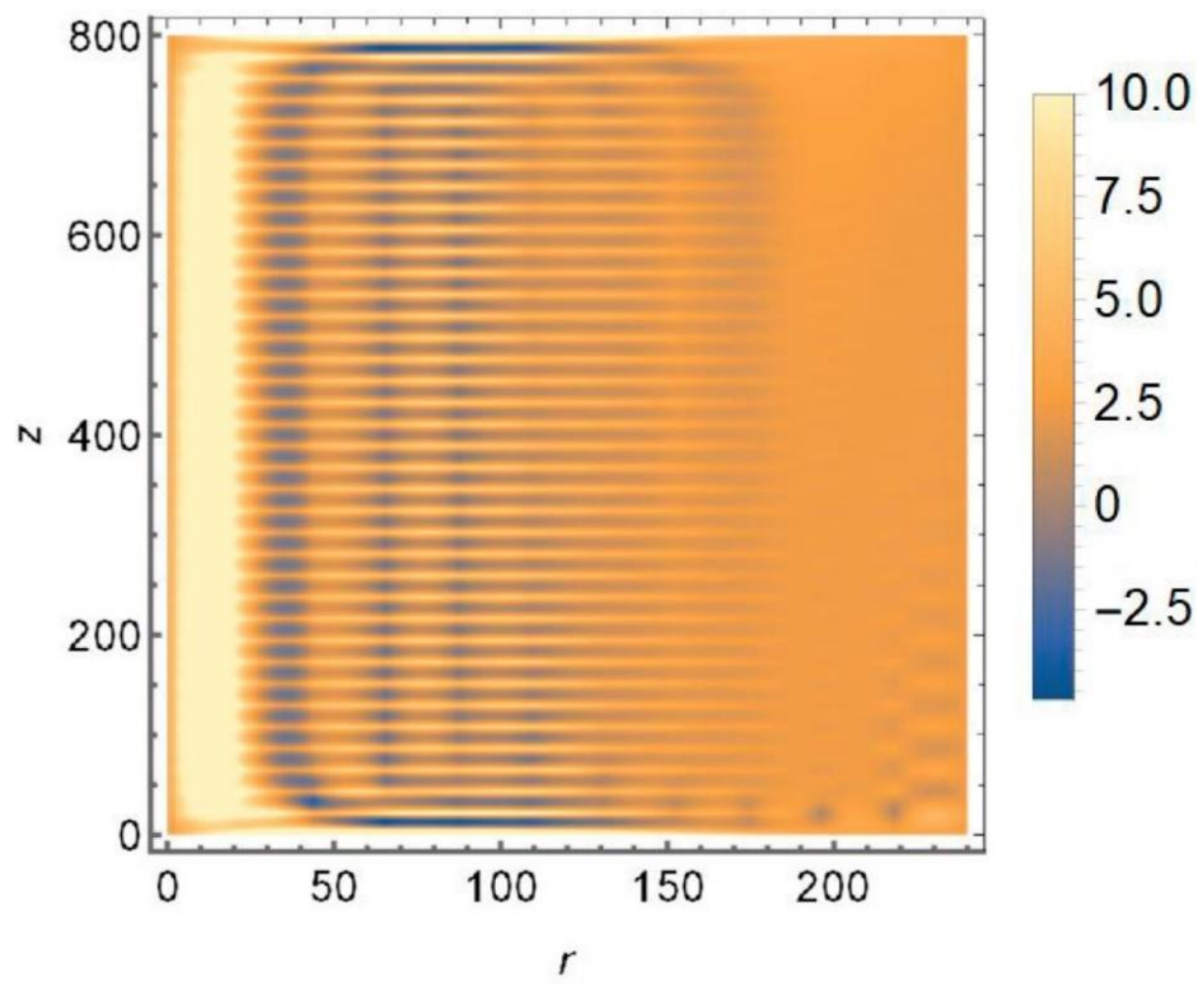

**FIG. 4.** The dependence of the azimuthal component of whirlwind velocity field on the coordinates $r$ and $z$ in meters for the function $f(z/L)$ from (43).

As we see from the structure of equation (21), arbitrary dependence on $z$ can be included in the Brunt–Väisälä frequency $\omega_g(z)$, and the solution for the stream function $\psi$ will remain the same. There arises the differential dependence of growth rate on height $z$: $\gamma \rightarrow \gamma(Z)$. The same change will occur for the expression of the vertical velocity component, but for the radial velocity component the following additional replacement will occur: $f'(Z) \rightarrow f'(Z) + f(Z)\gamma'(Z)t\coth[\gamma(Z)t]$. Again, we must solve the equation (19) numerically.

Let us formulate some concise summary of the simplest analytical model presented here. The simplest model can be applied to the initial stage of vortex generation, when all values differ little from equilibrium ones. To begin the process of tornado formation, an unstable stratified atmosphere with a presented parent thundercloud and an area with an initial swirl are presumed. At this stage the main movements are convective (poloidal). The model includes the initial $z$-elongation of the vortex (and changes $\frac{\partial}{\partial r} \gg \frac{\partial}{\partial z}$). We use incompressible flow ($\operatorname{div} v = 0$) and exclude viscosity (the model operates within the framework of ideal hydrodynamics). As a result, the law of conservation of mass and the law of conservation of angular momentum are satisfied for the vortex. Kinetic energy is not conserved, since the potential energy of the unstable atmosphere is converted into kinetic energy of motion. The hydrostatic balance is disrupted in the process of tornado generation.

Let us now pass on to the study of refined tornado models.

## 5. Tornado model with refined system of equations

Now let us consider the system of equation (18)-(20). In essence, the process in this case is rotating convection from the very beginning. Such a system can only be solved numerically by specifying the initial and boundary conditions. Let us recall that in ideal hydrodynamics we cannot set no-slip conditions for the tangential components of velocity at the boundary (but only non-passage conditions for the normal component of velocity). Therefore, we must have $v_z = 0$ at $z = 0, L$, and at $r = 0$ we must have $v_r = 0, v_\varphi = 0$. Let's choose $L = 2000$ m, $r_0 = 250$ m, and leave the remaining parameters (41) the same as in the previous Section. For the $\psi$ function, let's choose the corresponding solutions for the simplest model for all the selected parameter values as initial and boundary conditions (see (41) in the previous Section with $f(z/L)$ from (42)). In this case, calculations show that at the initial stage of the generation, the change in $\chi$ is rather small (~ 1-2 %) and its influence on the resulting solution is unnoticeable. At the initial stage, the influence of the azimuthal velocity component on the profile and the magnitude of the radial and vertical velocity components is very small. However, the azimuthal velocity component itself depends significantly on the initial and boundary conditions. Unlike the cases shown in Figs. 3 and 4, in this case a unidirectional differential rotation always occurs. However, the rotation can turn out to be local: only near certain regions along r and z. For example, such cases will be for the following initial and boundary conditions: 1) $v_\varphi = 0$ for $r = 0$, for z = 0, $L$, but $v_\varphi = v_{\varphi 0}$ at the periphery, or for 2) $v_\varphi = v_{r0}$, where $v_{r0}$, $v_{z0}$, and $v_{\varphi 0}$ are the solutions for the simplest model described in the previous Section. The case where the entire column L is involved in the rotation is shown in Fig. 5. Here we have chosen the following as the initial and boundary conditions: $v_\varphi = \sqrt{v_{r0}^2 + v_{z0}^2}$. We can also see that the vertical velocity component reaches a maximum value in the center of the whirlwind (upward flow) and radial inflow is maximal at $r_0$ near the earth surface. All components of the velocity tend to zero at the boundary of the whirlwind.

Let us recall that in the simplest model (Artekha, 2024; Onishchenko et al., 2023) the growth of the radial and vertical velocity components was approximately exponential, and the azimuthal component grew according to super exponential law (slowly at first and then extremely fast). For the refined model, all velocity components grow more consistently.

Latent heat plays the most important role in the intensification of a tornado: supercooled water vapor from the cloud descends near the axis of the tornado and condenses into droplets just in the eye wall. As a result, additional heat is generated, increasing the vertical flow in the eye wall. The influence of latent heat will change the dependence of the Brunt–Väisälä frequency on the coordinates. If we take into account not only the vertical, but also the radial dependence $\gamma(r, z)$, then the separation of variables cannot be applied and it is necessary to numerically solve the equation (20), and after then - Eq. (19).

It is known that the reliability of predicting the occurrence of tornadoes is extremely low - less than 25%. Therefore, the presence of torque and the fulfillment of energy conditions (for example, for a temperature gradient) are not enough for the spontaneous generation of a tornado. The numerical calculations performed show that the profile of the initial velocity field and its magnitude have a decisive influence on the possibility of a tornado occurring.

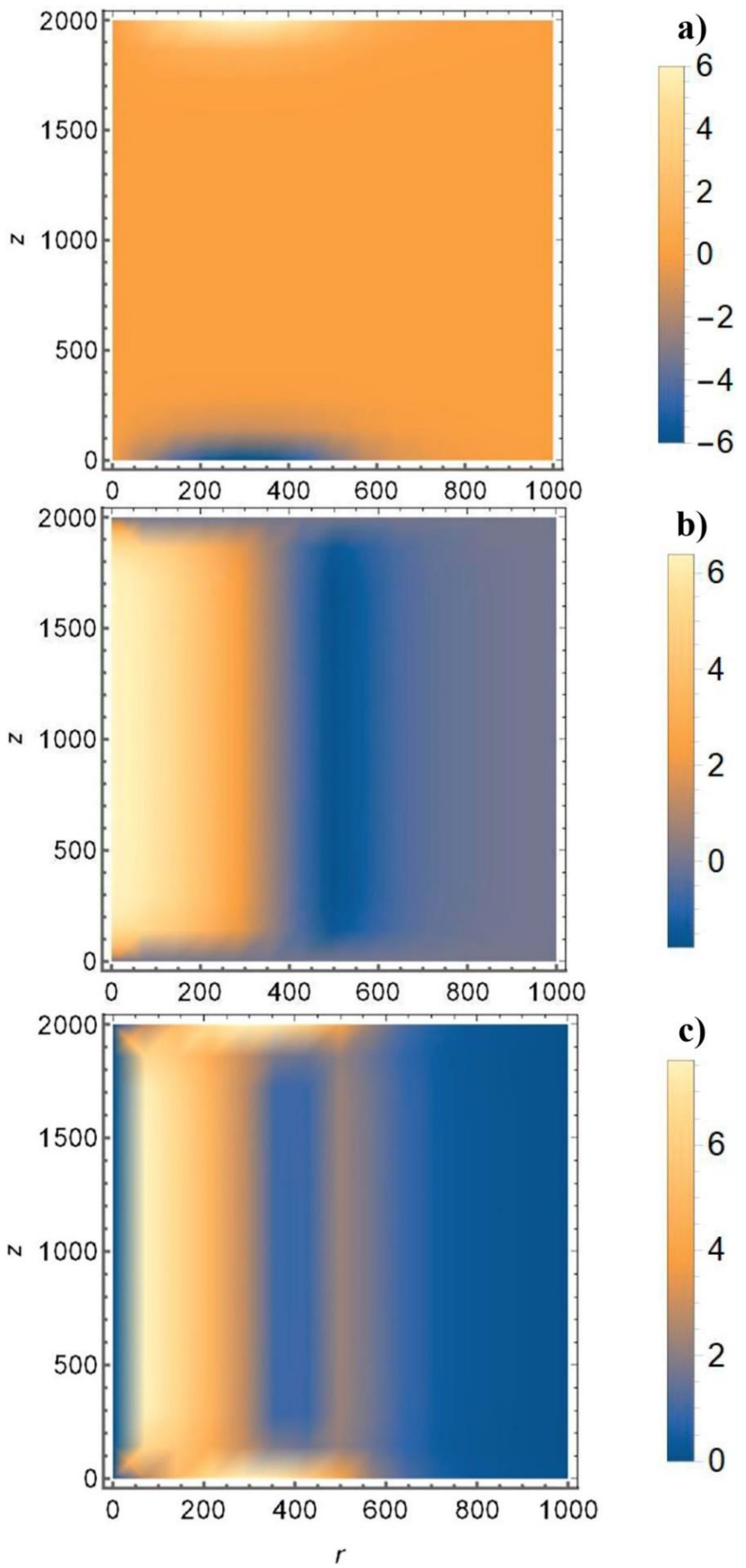

**FIG. 5.** The tornado velocity field in m/s depending on the coordinates $r$ and $z$ in meters for the refined model: a) $v_r(r,z)$; b) $v_z(r,z)$; c) $v_\varphi(r,z)$.

Let us formulate some concise summary of the refined model presented here. The refined model can be applied to the initial stage of tornado development, when the azimuthal motion becomes comparable to the poloidal motion. Unlike the simplest model, this model includes all nonlinear terms (any changes can be comparable). We use incompressible flow (div $v = 0, \rho \neq const$) and exclude viscosity (the model operates within the framework of ideal hydrodynamics). The law of conservation of mass and the law of conservation of angular momentum are satisfied for the tornado. Kinetic energy is not conserved: the potential energy of the unstable atmosphere is converted into kinetic energy of motion. The hydrostatic balance and the cyclostrophic balance are disrupted as tornadoes intensify.

Let us now move on to studying the complete system of equations.

**6. Tornado model with the complete system of equations**

Now let us consider the complete system of equation (13)-(15). We study the tornado generation process as rotating convection from the very beginning. The full system of equations can only be solved numerically by specifying all the initial and boundary conditions. In viscous case, we can set the following conditions: $v_r, v_\varphi, v_z = 0$ at $z = 0, L$, and at $r = 0$ we have $v_r = 0, v_\varphi = 0$. Let's choose again $L = 2000$ m, $r_0 = 250$ m, and leave the remaining parameters (41) the same as in two previous Sections. For the $\psi$ function itself and its derivatives, let's choose the corresponding solutions from the simplest model with all the selected parameter values as initial and boundary conditions (see (41) in Section 4 with $f(z/L)$ from (42)). As a result, all velocity components increase in a coordinated manner. The unidirectional differential rotation occurs for the entire column L. The vertical velocity component (upward flow) reaches a maximum value in the center of the whirlwind. The radial inflow is maximal at $r_0$ near the earth surface, and the radial outflow is maximal at $r_0$ near $z = L$.

All components of the velocity are regular on the axis of the vortex and tend to zero at the periphery. A whirlwind is generated and reaches a quasi-stationary stage within 15-20 minutes. The values of the velocity components correspond to the quantities observed in natural whirlwinds. The velocity profile (contour graphs of the velocity field on the $r - z$ plane) is shown in Fig. 6. We see the localization of the structure.

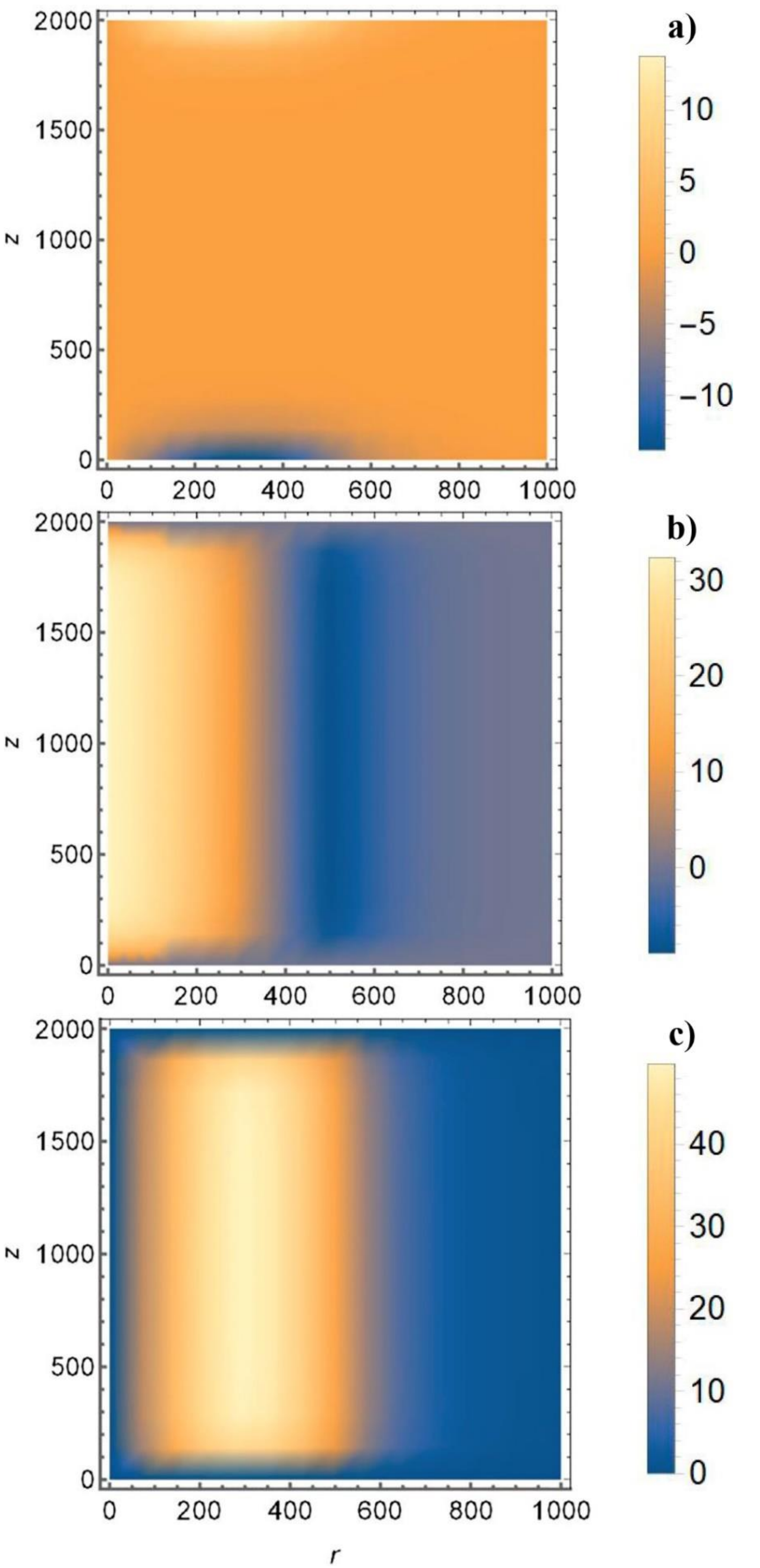

**FIG. 6.** The tornado velocity field in m/s depending on the coordinates $r$ and $z$ in meters for the mono-tornado in the framework of the full model: a) $v_r(r,z)$; b) $v_z(r,z)$; c) $v_\varphi(r,z)$.

Let us now consider the case where a tornado with two eyes is generated (Artekha, 2024). As parameters we will again use the values from (39), (40) and take $L = 2500$, $r_0 = 250$, $n = 2$. For the $\psi$ function itself and its derivatives, let's choose the corresponding solutions from the simplest model with all these selected parameter values (with $f(z/L)$ from (42)) as initial and boundary conditions (we will select the initial velocity component $v_{\varphi 0}$ from the linear case). The relative change in density $\chi(r, z)$ is presented in Fig. 7.

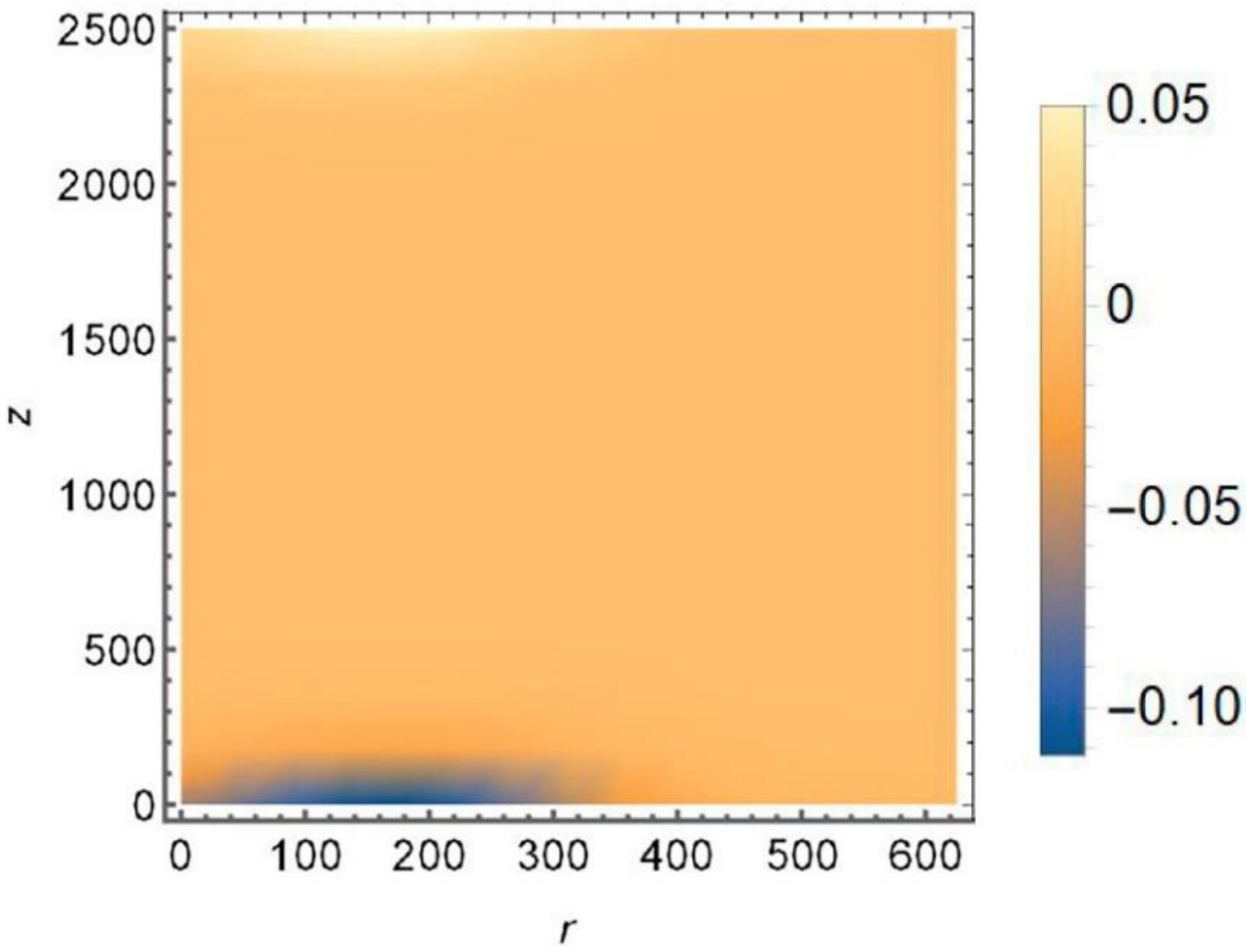

**FIG. 7.** The relative change in air density $\chi(r, z)$.

Computations show that all velocity components increase in a coordinated manner. The unidirectional differential rotation occurs for the entire column $L$ with two local maxima of $v_\varphi$ along $r$ (in such a case two eyes are detected). The vertical velocity component - downward flow - reaches a maximum value in the center of the tornado (at the axis), and the vertical velocity component - upward flow - reaches a maximum value at $z = \frac{L}{2}, r \sim 110$ m. The radial inflow is maximal at $r_0$ near the earth surface, and the radial outflow is maximal at $r_0$ near $z = L$, but there are also local extremums of $v_r$ at $r \sim 50$ m: outflow near the earth surface and inflow near $z = L$.

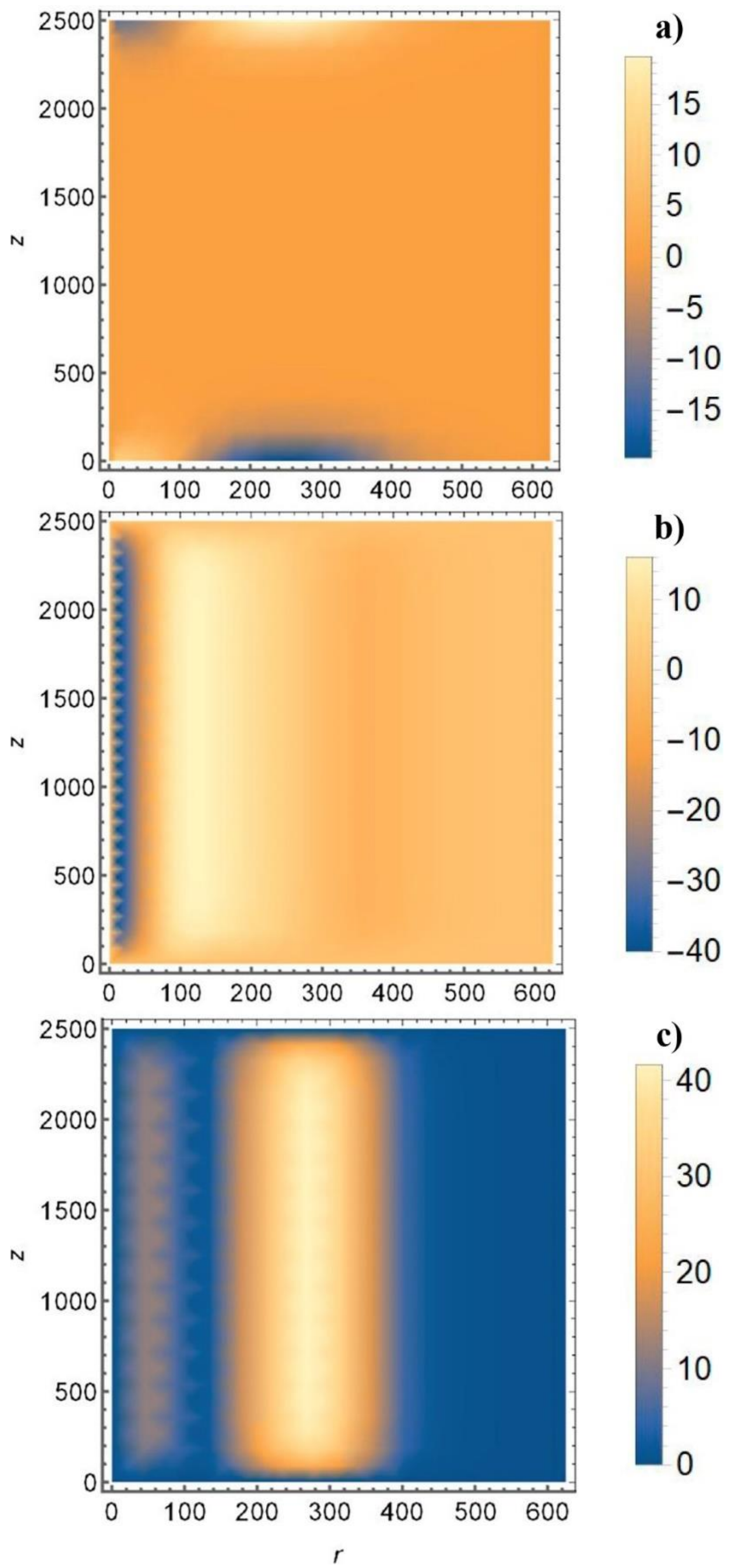

**FIG. 8.** The tornado velocity field in m/s depending on the coordinates $r$ and $z$ in meters for the tornado with two eyes in the framework of the full model: a) $v_r(r, z)$; b) $v_z(r, z)$; c) $v_\varphi(r, z)$.

All components of the velocity are regular on the axis of the tornado and tend to zero at the periphery. A tornado is generated and reaches a quasi-stationary stage within 10-30 minutes. In this case, it is possible to use both stationary final boundary conditions and slowly increasing boundary values (since in the process of self-organization the quantities can change). In the latter case, the boundary quantities change from background values to some final values: for example, by law of $\tanh(\gamma t)$. The final values of the velocity components correspond to the quantities observed in natural tornadoes. The velocity profile (contour graphs of the velocity field on the $r-z$ plane) is shown in Fig. 8. We see the localization of the vortex structure.

Let us formulate some concise summary of the full model presented here. The full model studies the tornado development as rotating convection from the very beginning to the quasi-stationary stage. The model includes all nonlinear terms (any changes arbitrary). We use incompressible flow (div $v = 0, \rho \neq const$). Unlike the previous two models, this model takes viscosity into account. The law of conservation of mass is satisfied for the tornado. Kinetic energy is not conserved: the potential energy of the unstable atmosphere is converted into kinetic energy of air motion. Unlike the previous two models, angular momentum can be not conserved due to the moment of frictional forces.

## 7. Discussion of results and conclusions

A possible mechanism of tornado generation (Artekha, 2024) can be associated with the Schwarzschild criterion for convective instability. At the preliminary stage, rising flows of moist air play a key role for tornadogenesis: they determine the formation of a cumulonimbus cloud itself and its maintenance. The process of vortex formation can proceed as follows: due to the super adiabatic temperature gradient, the overheated near-surface air mass (lighter) starts to rise, and the supercooled cloud-air mixture (heavier) starts to descend. Since the motion obeys the continuity equation, excess thermal energy leads to increased wind and intensification of the vortex. The highest velocities (Kosiba, Wurman, 2013; Kosiba, Wurman, 2023) are detected in the boundary layer (at $z \sim$ 30-150 m). Tornadoes often possess an eye-type structure (see, for example, (Bluestein, 2005; Wurman, Alexander, 2005), which can be described by two regions along $r$ as in (Onishchenko et al., 2023). However, tornadoes with two "eyes" (two maxima of the azimuthal velocity $v_\varphi(r)$ or of the wave structure of $v_\varphi(r)$) are also often observed (Wurman et al., 1996; Bluestein, Pazmany, 1999; Wurman, Gill, 2000; Kosiba et al., 2008; Wurman, Kosiba, 2013), which can be described by three regions along $r$ as in (Artekha, 2024).

In this work, a complete system of equations is derived, which includes an equation for the stream function, an equation for the azimuthal component of the velocity, and an equation for the change in density. The complete system is applicable for all stages of tornado development in the viscous case (full model). Some simplification of the system of equations for the case of ideal hydrodynamics was also carried out (but still considering a rotating convection). This system is applicable for the tornado growth stage (refined model). Finally, a separate equation for the stream function is obtained, applicable to the initial stage of vortex generation (simplest model). This is one of the possible methods for describing convective instability. All presented models describe the generation and growth of axially symmetric structures.

Within the framework of the simplest model, an analytical localized solution can be obtained by separating variables for the radial and vertical velocity components. Some initial twist plays an important role in the vortex generation: it symmetrizes poloidal movements about the tornado axis and breaks the symmetry with respect to the direction of rotation. The reasons for the initial rotation can be different (Artekha, 2024). Rotation with some definite profile increases due to the conservation of angular momentum (the law of conservation of mass and angular momentum is satisfied in this case). Due to the radial inflow, the tornado will intensify as it moves in the instability region with non-zero twist (non-zero helicity). The analytical tornado model proposed in (Artekha, 2024) includes several free parameters and is quite flexible. It allows us to describe a wide variety of tornadoes and whirlwinds observed in nature: their spatial structure, structure of the velocity field, values of extreme velocity components, their location etc. Some characteristic values of the simplest model are as follows: $r_0 = 50$m (range 30 m to 1.5 km), $L$ = 500 m (range 150 m to 1.5 km), $v_0 = 10$ m/s (range 5 m/s to 20 m/s), $v_{0\varphi} = 10$m/s (range 5 m/s to 30 m/s), $\gamma = 0.001 \div 0.01$ c$^{-1}$. The simplest model can be used at times $t \leq \frac{5}{\gamma} = 1000$ c. Mathematically, the solution can be combined like in a constructor: from blocks by radius and height. In this article we have constructed a solution from 8 linear dependencies by height (and 3 regions by radius). The resulting velocity field is shown in Fig. 2. In the lower part of the tornado, the radial flows converge to $R{\sim}120$ m (the maximal outflow near $R{\sim}60$ m, and the maximal inflow near $R \sim 250$ m). In the upper part of the tornado, inflows and outflows change places. The downward flow is maximal at the center of the tornado axis, the upward flow is maximal near $R{\sim}125$ m, $z = L/2$. At $R > 280$ m a downward flow is observed again. Tornado rotation is differential. The radial structure of the azimuthal velocity with two maxima is presented in Fig.1. The rate of increase of the azimuthal velocity component differs significantly from the rate of increase of the poloidal velocity components.

In the case of a nonlinear z-profile of the vertical flow, the equation for the azimuthal velocity component cannot be resolved by separation of variables. This article examines two cases where many regions with opposite rotation directions arise (see numerical results in Figs. 3 and 4). To obtain a correct result in such cases, it is necessary to take viscosity into account.

Within the framework of the refined model, some unidirectional differential rotation occurs. However, the value of the azimuthal velocity component depends on the initial and boundary conditions. For instance, the rotation can be local, only near some regions (r, z). An example of the single whirlwind with rotation of the entire column L is shown in Fig. 5. In this case, all velocity components grow rather consistently.

Within the framework of the full model, the tornado development is considered as rotating convection from the very beginning to the quasi-stationary stage. All velocity components change (increase) in a coordinated manner. The excess energy in the viscous case can turn into kinetic energy of rotation and increase the angular momentum ("heat engine"). The unidirectional differential rotation always occurs. The velocity component profiles are smoothed. This article presents two cases (results of numerical calculations) depending on initial and boundary conditions, namely, a single whirlwind and a two-eyed tornado: dependences with two maxima of $v_\varphi$ in *r*-profile. The corresponding velocity fields are shown in Figs. 6 and 8.

Thus, all proposed models are based on the already existing convective instability, and each model (in its area of applicability) is capable of describing rather adequately the main characteristics of a tornado.

If, with all the previously selected boundary conditions, we change only the initial conditions in the main air mass, then with a delay of only a few seconds the same tornado wind flows will be established as before (they are shown, for example, in Fig.8). If we also change the boundary conditions (even significantly), but make them change over time, gradually moving to the previous boundary conditions, then the solution follows these changes and with a delay of tens of seconds moves to the previous solution (shown, for example, in Fig.8). The following multiplicative coefficients were chosen as some disturbance:

$$1 + kF_1\left[\frac{2\pi n_1 z}{L}\right]F_2\left[\frac{2\pi n_2 r}{q_1 r_0}\right](1 - \tanh(\gamma_t t)),$$

with some arbitrary values of $0 < k < 0.9, \gamma_t$; $n_1, n_2 \in N$, and various degrees of sines or cosines were chosen in an arbitrary order as functions $F_1$ and $F_2$.

There is a wide variety of tornado shapes in nature, so it is worth explaining why we chose the cylindrical shape (“stovepipe”). The only reason is that for the simplest model we can then obtain a rigorous analytical solution to the problem. After this, we can compare with this analytical solution: 1) the numerical solutions of the refined model within the framework of ideal hydrodynamics, and 2) the numerical solution in the viscous case. The solutions turned out to be close, which means that these models can be trusted within the limits of their applicability (any presented figure is for illustrative purposes only and shows one of many possible tornado variants). The numerical solution of all three models is not tied in any way to the shape of the tornado (the shape can be set arbitrarily by boundaries).

There are some difficulties along the path of a direct quantitative comparison of different model descriptions and the real phenomenon. Any direct measurements of characteristics inside a tornado are impossible due to the destruction of measuring devices. There are no direct measurements of the density field, temperature field, or velocity field covering the entire tornado region (all data are estimates). And without such data, quantitative comparison with the real phenomenon is difficult. Some of the difficulties of the radar sounding of tornadoes and data interpretation are described in more detail in (Artekha, 2024). Any numerical calculations also use different hypotheses of tornado formation and assumed (not known) initial and boundary conditions. Other difficulties that stand in the way of accurately describing real tornadoes are discussed later in this Section. Some comparative analysis of existing tornado models is presented in (Kim, Matsui, 2017). Due to the wide variety of manifestations of tornadoes and whirlwinds, any comparisons of computations with reality (with one specific tornado) or with other models are possible only if all specific conditions are specified or could only be made qualitatively (by form or by order of magnitudes). The simplest model with an analytical solution (Artekha, 2024) allows one to describe tornadoes with a wide variety of characteristics by choosing free parameters, and the refined and full models can then obtain similar characteristics of a tornado by choosing initial and boundary conditions. Previously, in the paper (Artekha, 2024) a comparison of the results of the simple analytical model with some field observations of tornadoes was presented. So, for example, all the structures shown in Fig. 9 of the article (Artekha, 2024) and their characteristics are similar to the corresponding values presented in Fig. 1 of the article (Kosiba et al., 2008) and Fig. 7 of the article (Wurman, Alexander, 2005), if we take into account the violation of axial symmetry due to tornado translational motion; The three-dimensional structure of the movements in Fig. 10 of the article (Artekha, 2024) and the order of all quantities resemble some observations of tornadoes from (Kosiba et al., 2008; Kosiba, Wurman, 2010; Lee, Wurman, 2005). It is only necessary to take into account that in all the models presented here the flows either are closed or pass into background values at the periphery, but radar observations provide data only on a part of the region occupied by a tornado (limited by $r < r_{max}$ , $z_{min} < z < z_{max}$), and the speeds at such the boundaries remain quite significant. The heterogeneities of tornado winds are associated with heterogeneities in background values and the terrain. Similar to (Onishchenko et al., 2023; Artekha, 2024), we choose $\delta = 0.08, r_1 \approx 1.3121, m \approx 1.172114$ and obtain the normalized tangential velocity. The results of the analytical model for the dimensionless tangential velocity, presented here in Fig. 9 by the black solid curve, fit neatly among the measured data from the paper (Kosiba, Wurman, 2010) for the 30 May 1998 Spencer, South Dakota, tornado. The results of the simplest analytical model, presented here in Fig. 10 by the black solid curve, fit neatly among the measured data from the paper (Lee, Wurman, 2005) for the Mulhall Tornado on 3 May 1999.

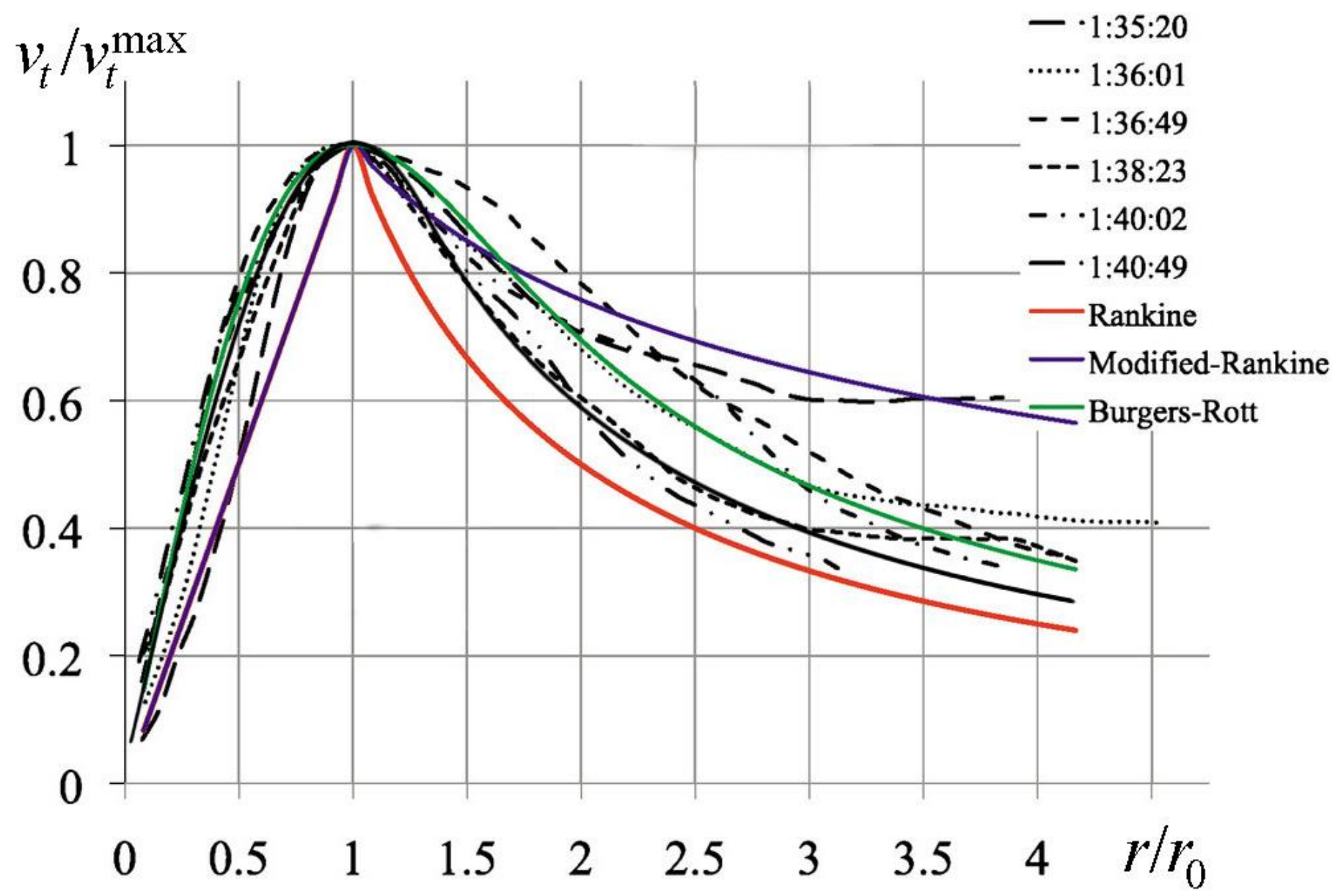

**FIG. 9.** Normalized radial profiles of the axisymmetric tangential winds at a height of 40 m AGL from 0133:37 to 0140:49 UTC for the 30 May 1998 Spencer, South Dakota, tornado. Results of the simplest analytical model are shown by the black solid line.

We will notice that purely analytical solutions of the simplest model have limitations: their extrapolation to large times can give incorrect results, since observations show that the radius of maximum winds and the vertical profile of the tornado velocity field change with time, i.e., it should be $r_0 \to r_0(t), f(z) \to f(t,z)$. There are no such restrictions in the numerical calculation for the refined and complete models (but the refined model does not smooth velocity profiles).

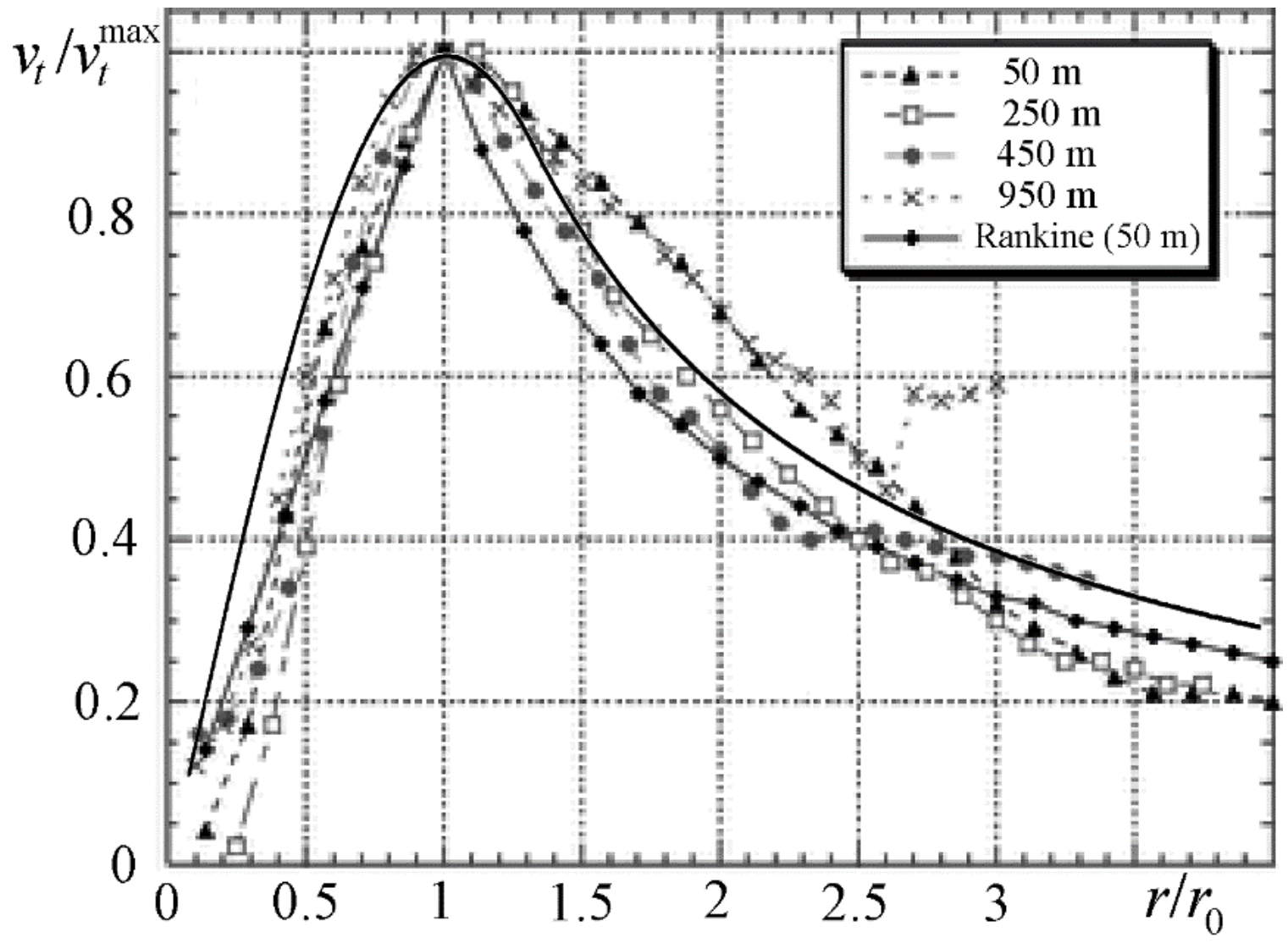

**FIG. 10.** The normalized radial profiles of the axisymmetric tangential winds of the Mulhall tornado at 0310:03 UTC 4 May 1999 at altitudes of 50 m, 250 m, 450 m, and 950 m. Results of the simplest analytical model are shown by the black solid line.

The visible shape of a tornado is determined by flying debris (impurities). And since the azimuthal velocity reaches its maximum at a certain height $h_0$, the centrifugal force $m_0 v_\varphi^2/r$ increases from the ground up to this height $h_0$. As a result, in any tornado, the flying debris forms a truncated cone, expanding from the ground surface. Tornadoes in the shape of a truncated cone up to the parent cloud ("wedges") are often encountered (but not always the case). Not all field data or high-resolution radar retrievals detect a cone (on the contour plot $r-z$ this is the shape of a "nose"). For example, Fig. 6 (a)-(e) from (Kosiba, Wurman, 2010) demonstrate such a "nose", but Fig. 6 (f)-(h) there, on the contrary, shows a "stovepipe", just like Figs. 4 (a), 9 (a), (c), (e) from (Lee, Wurman, 2005). It will not be difficult to specify any form, including a cone (through boundary conditions), in the refined and full models. For example, if we take as boundary conditions the solution of the simplest model with the parameters $\delta = 0.08, r_1 \approx 1.3121, m \approx 1.172114, n = 0.75$, but the value of $r_0(z)$, increasing with the growth of $z$, then we obtain a tornado in the shape of a wedge (see Fig. 11). The maximum winds are rather close to the peak values measured by the radar (see Fig. 6 (a) in (Kosiba, Wurman, 2010)). Of course, the figure is only illustrative (one cell only), since the exact boundary conditions are unknown, and some issues must be taken into account and solved (see below).

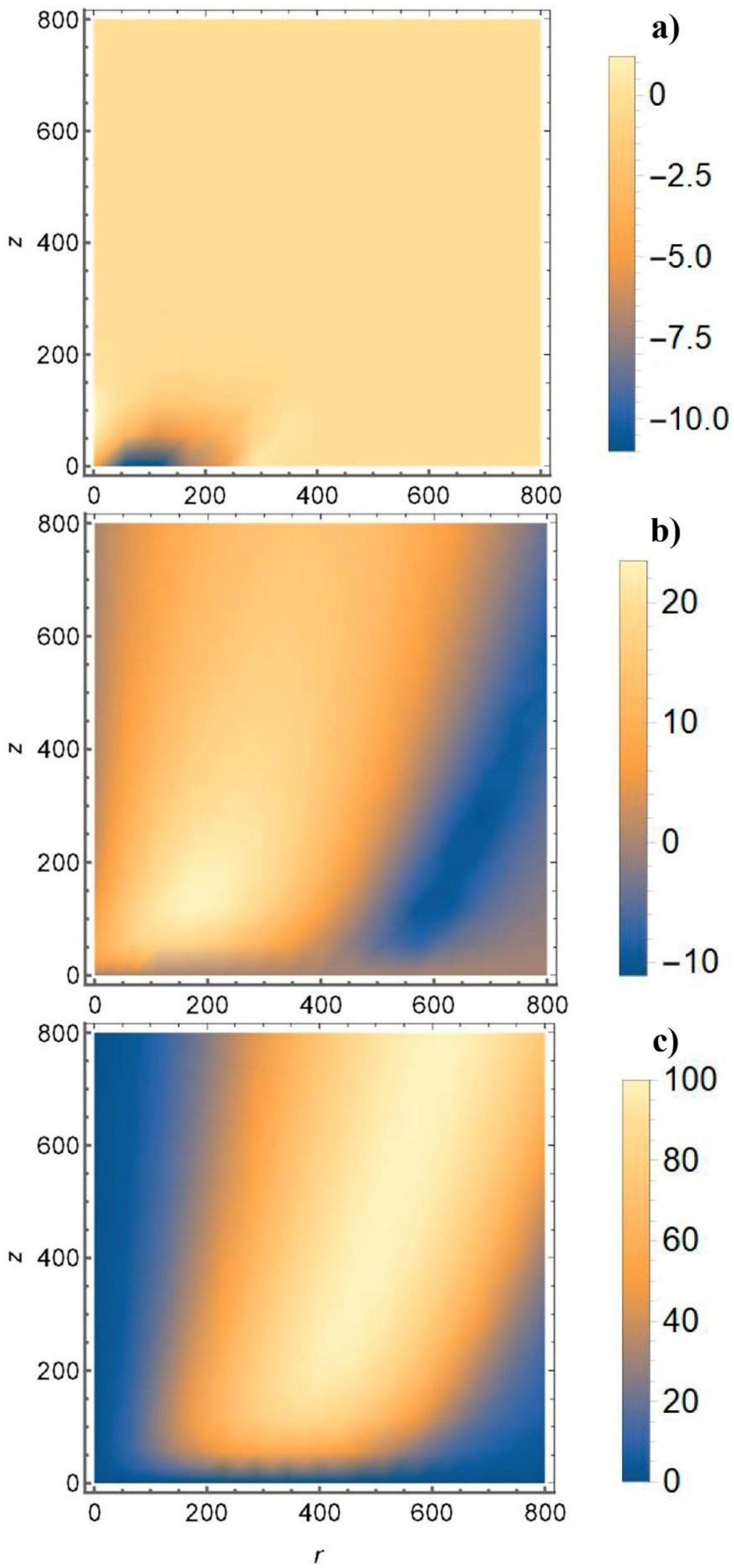

**FIG. 11.** The tornado velocity field in m/s depending on the coordinates $r$ and $z$ in meters for the one-cell wedge tornado (in the framework of the full model): a) $v_r(r,z)$; b) $v_z(r,z)$; c) $v_\varphi(r,z)$.

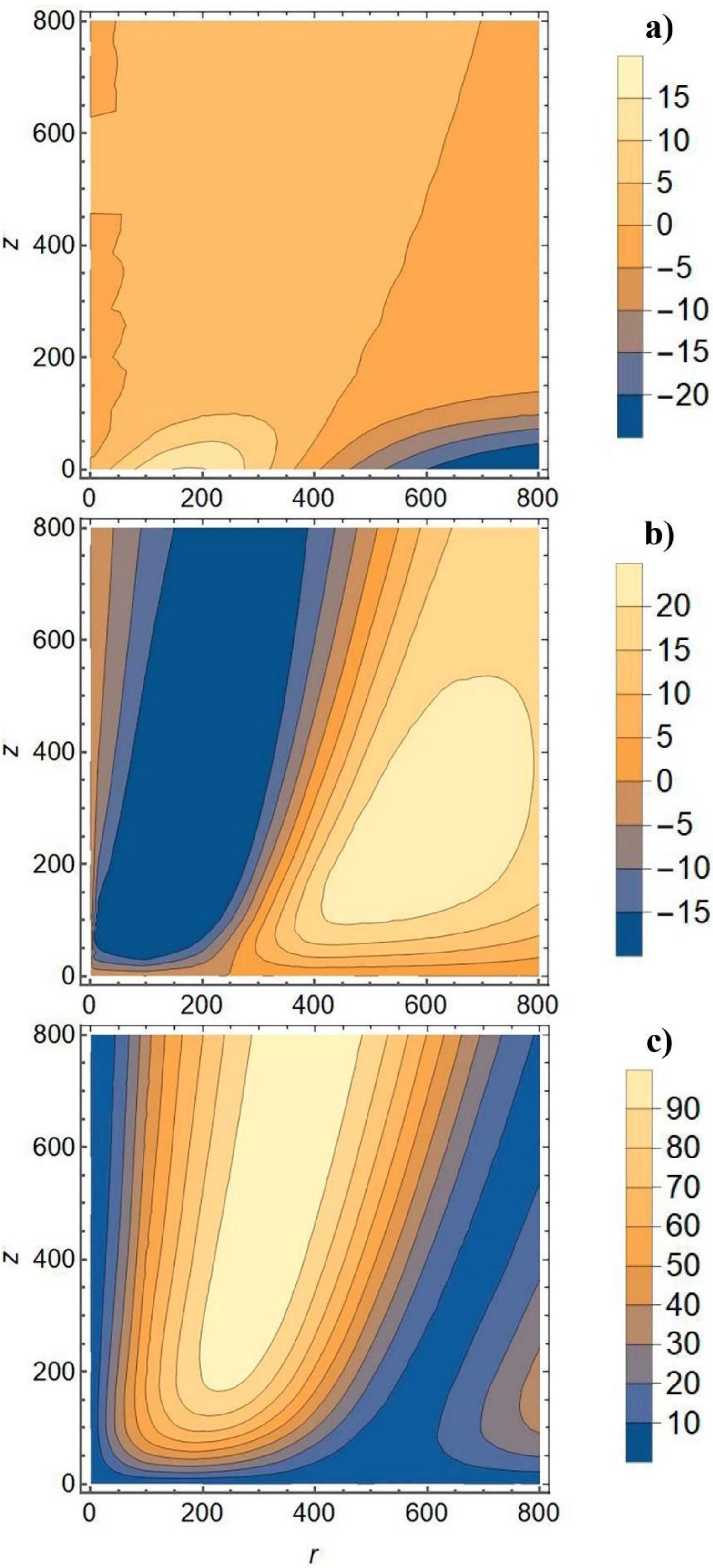

**FIG. 12.** The tornado velocity field in m/s depending on the coordinates $r$ and $z$ in meters for the two-cell wedge tornado (the full model): a) $v_r(r,z)$; b) $v_z(r,z)$; c) $v_\varphi(r,z)$.

If we take as boundary conditions the solution of the simplest model with the parameters (39), (40), $L = 2000$ m, $r_0(0) = 700$ m, $n = 0.75$, and the value of $r_0(z)$, increasing with the growth of $z$, then we obtain a two-cell tornado in the shape of a wedge (see Fig. 12). The maximum winds are rather close to the peak values measured by the radar (see Fig. 6 (a)-(d) in (Kosiba, Wurman, 2010)). The obtained structure of movements also has common features with the measured data: maximum inflows-outflows are observed near the earth surface (we see outflow from $r = 0$ to 300 m, inflow – from $r = 400$ m and further); up to $r = 200$ m, we see downward movement, and from about 300 m – upward movement; maximum azimuthal velocities are observed about $r \sim 200$-300 m. Of course, the figure is also illustrative, since the exact boundary conditions are unknown (heterogeneities of tornado velocity field are associated with heterogeneities in background values and the terrain, but we took smooth functions). In addition, some problems must be taken into account and resolved (see discussion below).

Thus, the results presented in this article and in the article (Artekha, 2024) suggest that in order to improve tornado occurrence forecasts (including using remote sensing), it is necessary to test the following conditions:

1) Several quantities had to exceed threshold values (necessary conditions). For example, the surface layer temperature must be high enough (heat engine works with a super adiabatic temperature gradient):

$$\frac{dT}{dz} > \frac{\gamma_a - 1}{\gamma_a} \frac{g\rho_0 T}{P_0}, \tag{44}$$

and the greater the inequality, the greater the reserve of thermal energy. At the same time, near the axis of rotation in the cloud, there should be a denser air-cloud mixture (most likely of lower temperature).

2) Size of the unstable area with some nonzero angular momentum (rotating convection) must be large enough for a tornado to have time to organize $\Delta r \gg 2r_0$.

3) Since boundary conditions play a key role in establishing tornado wind flows, especially the conditions at the top of the tornado – in the parent cloud, then the proximity of the spatial profile (in form) of the motion inside the cloud to the solution found is important. Only then these motions "picked up" by the entire air mass below and self-consistently amplified. Therefore, it is necessary to subtract the average translational velocity of the cloud and determine the radial profile of the azimuthal velocity component $v_\varphi(r)$. Next, it is necessary to compare this profile with the profile of the azimuthal velocity component $v_\varphi^{torn}(r)$, which is typical for a tornado (see results of this article and Artekha, 2024; Onishchenko et al., 2023). It can be assumed that another necessary condition is the small difference between these quantities, for example:

$$\frac{1}{r_2}\int_0^{r_2}\frac{|v_\varphi^{torn}(r)-v_\varphi(r)|\,dr}{v_\varphi^{torn}(r)} < k_\varphi \ , \tag{45}$$

where $k_\varphi \sim 0.2$, and we compare the form in the main area $(0, r_2)$ of possible tornadoes (the periphery only affects accompanying vortices or multi-tornadoes). As soon as the radial inflow to the denser mass near the axis begins, the formation of the tornado begins. Apparently, at the moment of the start of the generation, the translational velocity should not be very high so that the tornado structure is established along the entire length of the column (once the structure has formed, the forward speed has little effect on the stability of the tornado).

Now we discuss some of the prospects for more comprehensive storm-scale modeling. Let us start with some minor comments. Air humidity can be included in the equilibrium density $\rho_0 = P_0\mu/T$ through the change in molar mass: $\mu = (1-q)\mu_0 + q\mu_1$ , where $\mu_0 \approx 0.029$ kg/mol is the molar mass of dry air, $\mu_1 \approx 0.018$ kg/mol is the molar mass of water, $q$ is specific humidity. As humidity increases, air density decreases monotonically. For example, at a temperature of 30 °C and normal pressure, the density of saturated moist air is approximately 1.2 % less than the density of dry air, which significantly affects the intensification of convection. The onset of precipitation along the axis can occasionally trigger the beginning of the tornado's trunk to descend.

However, the most important thing is incorporating feedback on latent heat. Inflow of moist air from the outside to the tornado wall and outflow of supercooled moist air, descending from the cloud, from the tornado axis to the tornado wall causes condensation into droplets just in the eye wall. As a result, additional heat is generated in the tornado wall region near the earth surface ($z$ < 100 m). The presence of large amounts of (charged) dust and ions in this turbulent area accelerates moisture condensation. More precisely, this region can be approximated in form by the earlier introduced function $f'(z)$ with respect to $z$, and by the function $v_\varphi(r)$ with respect to $r$, i.e. $F_{reg} = f'(z)v_\varphi(r)$. In essence, the release of latent heat is equivalent to an increase in the temperature gradient. As a first approximation, for this unstable region, we can make the following replacement:

$$\omega_g^2 \equiv -\gamma^2 \to -\gamma^2 + \frac{L_V}{c_p}\frac{g}{T}\frac{\partial q}{\partial z}, \tag{46}$$

where $L_V \approx 2555$ kJ/kg is the specific heat of condensation; $c_p \approx 1005$ J/(kg · K) is the specific heat capacity of dry air at constant pressure; $\frac{\partial q}{\partial z} < 0$ is the vertical gradient of specific humidity. As a first approximation, this decrease can be taken as constant: $\frac{\partial q}{\partial z} \sim -4 \cdot 10^{-5}$ m$^{-1}$. However, it seems physically more correct to consider that the greatest condensation (the greatest release of latent heat) in the inflow region occurs where the greatest centrifugal acceleration is. Therefore, the new addition in (46) must be multiplied by the normalized function $F_c(r,z)/F_{cmax}$, where $F_c(r,z) = f'(z)v_\varphi^2(r)/r$. Thus, it is possible to take this effect (dependence of $\gamma(r,z)$) into account in the numerical calculation.

A more complex issue is related to taking viscosity into account. Since the main objectives of this work were to further develop an analytical model of tornado (Artekha, 2024), to compare the results of the simplest, refined and complete tornado models for the generation stage, growth and extrapolation to saturation, then viscosity was implemented as a constant coefficient in the direct numerical calculation. The calculations demonstrate similar results. However, since it is impossible to take into account all the scales of vortices (down to millimeters) in a direct numerical calculation, it is necessary to use turbulent viscosity, which depends both on the scales of the grid and on the characteristics of the turbulent flow itself, i.e. it is a variable value. The order of magnitude of turbulent viscosity can be estimated as $\nu_t \sim 0.1 v_\varphi r_0$. Turbulent viscosity can differ from molecular viscosity by $\sim 10^2 - 10^6$ times depending on different approaches and grid step (as the step tends to zero, the turbulent additive also tends to zero, and only the usual molecular viscosity remains). To take into account the influence of turbulence on the mean motions, it is necessary to close the turbulent equations and to find the turbulent viscosity (for example, there are the simplest models by Boussinesq or Prandtl). From the author's viewpoint, the following approaches are the most promising for more comprehensive storm-scale modeling: 1) LES – Large Eddy Simulation (Sagaut, 2006) with subgrid-scale modelling (for example, Smagorinsky's formula), or 2) RSM – the Reynolds stress equation model, which allows taking into account high degrees of anisotropy, significant streamline curvature, rotational effects (Pope, 2000; Launder, Hanjalic, 2011). To further refine the results of the article, the author plans to use these two approaches in future work.

Another difficult issue is related to taking into account non-axisymmetric disturbances. Taking such influences into account is especially important when passing through complex terrain, strong urban developments, or during the period of weakening of a tornado (to describe its self-destruction). Nonlinear modal decomposition methods and model reduction can be used. The solution scheme is as follows.For each velocity component, we seek a solution in the form of a superposition of the basic axisymmetric flow ($m = 0$) and a finite set of selected non-axisymmetric modes ($m = 1, 2, 3, \ldots$):

$$v_i(t, r, \varphi, z) = v_i^0(t, r, z) + \sum_m \left( v_i^m(t, r, z) e^{i(m\varphi - \omega t)} + v_i^{-m}(t, r, z) e^{-i(m\varphi - \omega t)} \right).$$

Next, the resulting system of nonlinear equations for the modes is solved numerically. The integer azimuthal number $m = 0$ is responsible for the axisymmetry (contributes to the change in the basic axisymmetric solution), $|m| = 1$ is responsible for the offset / tilt, $|m| = 2$ determines the ellipticity, etc. The imaginary part of the cyclic frequency $\omega$ determines the increment (decrement). When modes $m_1$ and $m_2$ interact, modes $|m_1 \pm m_2|$ can arise.

The biggest challenge in modeling real tornadoes is accounting for the influence of the entrained passive admixture (particles, dust, debris) on the characteristics of the tornado (Lewellen et al., 2008; Baker, Sterling, 2017) and on the flow turbulence. Due to different aerodynamic drag (see, for example, (Zastawny et al., 2012)), particles of different masses and different shapes have different distributions along the height of the vortex – a complex stratification occurs (Wang, Maxey, 1993). The effective air density also changes. These modifications are easy to take into account. From general considerations it is clear that impurity particles (heavier than air) tend to reduce velocities inside the vortex and increase the radius of the vortex. They increase the inertia of the vortex in relation to any change in radial and azimuthal flows, resist an increase in the speed of the vertical flow, and when the lifting force decreases, they begin to fall down, entraining air and contributing to the weakening (and destruction) of the tornado. However, for a more accurate calculation it is necessary to use the equations of multiphase media (Crowe et al., 2011), to include the mutual influence of air flows and particles, taking into account their different masses, sizes, and shapes. Small particles with high inertia (of a "bulk" shape) can effectively dampen small-scale gas turbulence, while large particles (angular, elongated, and flat shapes) can generate turbulence due to vortex wakes. Therefore, modification of turbulent viscosity is required. All these questions do not yet have any complete solutions.

Let us now discuss the concept of the swirl ratio $S$ for the vortex. This concept is widely used in laboratory modeling of vortices (including tornado-like ones), that is, under completely controlled conditions. The given article presents the results of the simplest analytical model, the results of direct numerical calculations for the refined model and for the full model (with constant turbulent viscosity). The result depends on the selected parameters, or initial and boundary conditions, respectively. The simplest analytical model is applicable to the initial stage of vortex generation, when the maximum tangential velocity $v_t \ll v_z$ (the graphs shown are the result of extrapolation to longer times); therefore, in the area of applicability, the swirl ratio $S \ll 0.5$ is always. The refined model is applicable to the initial stage of tornado growth ($v_t^{max} < v_z^{max}$; the graphs shown are the result of extrapolation to some longer times also); in the area of applicability, the swirl ratio $S < 1$. According to (Kosiba, Wurman, 2010), the value of $S \sim 1$ corresponds to a two-cell vortex. With the selected initial and boundary conditions we have chosen for the full model, we obtain $S > 1$. However, it is worth remembering that this parameter $S$ does not guarantee the presence of a single vortex or, conversely, multiple vortices, but only indicates a more probable trend. In reality, tornadoes with a wide variety of characteristics are observed, including $S > 1$ (see, for example, (Kosiba, Wurman, 2010) for the Spencer tornado). There is an opinion that this concept is too vague and of little use for a real non-stationary tornado (Nolan, 2006). However, recently important attempts have been made to bring order to this concept and its application (Wang et al., 2022). Future research will show whether the swirl ratio can play a predictive role for tornadoes, or not.

The tornado phenomenon is very complicated, so it is impossible to reveal all the connected nuances of this extensive research topic in one paper. Some important further direction in the development of the suggested tornado models is the search for quantitative relationships of found parameters with the observed characteristics of an unstable atmosphere and a cumulonimbus cloud. The most important direction is the statistical analysis of the fulfillment of the conditions for the formation of a tornado and the refinement of the proposed criteria. Involving artificial intelligence (or machine learning) in the process of real-time remote sensing could help in this matter.

**CRediT authorship contribution statement**

**S.N. Artekha:** Conceptualization, Writing – original draft, Visualization, Validation, Software, Methodology, Investigation, Formal analysis.

**Declaration of competing interest**

The authors declare that they have no known competing financial interests or personal relationships that could have appeared to influence the work reported in this paper.

**Acknowledgments**

The work was carried out within the framework of the State assignment on the topic of fundamental scientific research “Monitoring” IKI RAS (122042500031-8).

**Data availability**

Data will be made available on request.